\documentclass[a4paper, 11pt]{article}
\usepackage[utf8]{inputenc}
\usepackage[english]{babel}
\usepackage{geometry}

\usepackage{jcappub}

\usepackage{graphicx}
\usepackage{wrapfig}
\usepackage{caption}
\usepackage{subcaption}
\usepackage{afterpage}

\usepackage{physics}
\usepackage{amssymb}
\usepackage{amsfonts}
\usepackage{tensor}
\usepackage{siunitx}

\usepackage{enumitem}
\setlist[description]{labelindent=0em,leftmargin=0em}

\makeatletter\g@addto@macro\bfseries{\boldmath}\makeatother%

\newcommand{\0}{\nonumber}
\newcommand{\iu}{\mathrm{i}\mkern1mu}
\DeclareMathOperator\cst{cst}

\allowdisplaybreaks

\title{A novel family of rotating black hole mimickers}

\author{Jacopo Mazza,}
\author{Edgardo Franzin,}
\author{and Stefano Liberati}

\affiliation{SISSA, International School for Advanced Studies,\\via Bonomea 265, 34136 Trieste, Italy}
\affiliation{INFN, Sezione di Trieste,\\via Valerio 2, 34127 Trieste, Italy}
\affiliation{IFPU, Institute for Fundamental Physics of the Universe,\\via Beirut 2, 34014 Trieste, Italy}

\emailAdd{jmazza@sissa.it}
\emailAdd{efranzin@sissa.it}
\emailAdd{liberati@sissa.it}

\abstract{The recent opening of gravitational wave astronomy has shifted the debate about black hole mimickers from a purely theoretical arena to a phenomenological one. In this respect, missing a definitive quantum gravity theory, the possibility to have simple, meta-geometries describing in a compact way alternative phenomenologically viable scenarios is potentially very appealing. A recently proposed metric by Simpson and Visser is exactly an example of such meta-geometry describing, for different values of a single parameter, different non-rotating black hole mimickers.
Here, we employ the Newman--Janis procedure to construct a rotating generalisation of such geometry. We obtain a stationary, axially symmetric metric that depends on mass, spin and an additional real parameter $\ell$. According to the value of such parameter, the metric may represent a rotating traversable wormhole, a rotating regular black hole with one or two horizons, or three more limiting cases. By studying the internal and external rich structure of such solutions, we show that the obtained metric describes a family of interesting and simple regular geometries providing viable Kerr black hole mimickers for future phenomenological studies.}


\arxivnumber{2102.01105}

\begin{document}

\maketitle \flushbottom 


\section{Introduction} \label{sec:introduction}

Spacetime singularities are a generic prediction of General Relativity (GR) \cite{hawking_large_1973}, under a set of physically reasonable (yet fairly stringent) assumptions. They represent points at which geodesics terminate abruptly and, for this reason, their appearance is usually interpreted as marking the breakdown of the theory.

It is commonly believed that quantum gravitational effects will prevent the formation of singularities. A complete description of such small-scale effects is still out of reach; however, one might wonder whether they propagate to larger scales, at which gravity is well described within the framework of differential geometry. By studying the phenomenology of suitably constructed effective geometries  ---  hopes are  ---  one may get insight into the process whereby the singularity is regularised and, therefore, also into the quantum gravity theory responsible for it \cite{carballo-rubio_phenomenological_2018}.

An agnostic analysis \cite{carballo-rubio_opening_2020,carballo-rubio_geodesically_2020}, almost entirely based on geometric arguments, shows that the spectrum of qualitatively different regular geometries is not as wide as one might expect. One alternative emerging from such classification, and particularly relevant to the present discussion, is represented by wormholes: heuristically, the throat of the wormhole, reached at finite affine distance, defocuses the null geodesics that would otherwise converge (and end) at the singularity. Such defocusing point may lie in an untrapped region, or be cloaked by an event horizon: in the former case, the wormhole is traversable (in the sense of Morris and Thorne \cite{morris_wormholes_1988}); in the latter, the spacetime contains a regular black hole.

Simpson and Visser \cite{simpson_black-bounce_2019, simpson_vaidya_2019, lobo_novel_2020} recently proposed a static and spherically symmetric metric that smoothly interpolates between these possibilities. The line element is (henceforth: SV metric):
\begin{equation}
     \dd s^2 = - \left(1- \frac{2M}{\sqrt{r^2 + \ell^2}} \right) \dd t^2 + \left(1- \frac{2M}{\sqrt{r^2 + \ell^2}}\right)^{-1}\!\dd r^2 + (r^2+\ell^2) \left [\dd \theta^2 + \sin^2 \theta \dd \phi^2 \right ],
     \label{eq:SV}
\end{equation}
where $M\geq 0$ represents the ADM mass and $\ell>0$ is a parameter responsible for the regularisation of the central singularity\footnote{In \cite{simpson_black-bounce_2019} $a$ was used in lieu of $\ell$. However, $a$ is widely used as the spin parameter for rotating spacetimes, thus we renamed the regularisation length scale $\ell$.}; note that $r \in (-\infty, + \infty)$. Despite its simplicity   ---   it is a minimal modification of the Schwarzschild solution, to which it indeed reduces for $\ell=0$   ---   this metric is remarkably rich: trimming the value of $\ell$, it may represent
\begin{itemize}
    \item a two-way, traversable wormhole à la Morris-Thorne for $\ell>2M$,
    \item a one-way wormhole with a null throat for $\ell = 2M$, and
    \item a regular black hole, in which the singularity is replaced by a bounce to a different universe, when $\ell<2M$; the bounce happens through a spacelike throat shielded by an event horizon and is hence dubbed ``black-bounce'' in \cite{simpson_black-bounce_2019} or ``hidden wormhole'' as per~\cite{carballo-rubio_geodesically_2020}. 
\end{itemize}

Such variety is particularly appealing. Regular black holes and traversable wormholes represent morally distinct scenarios, and have been studied thoroughly, though separately: see for instance \cite{bardeen_non-singular_1968,  dymnikova_vacuum_1992, ayon-beato_bardeen_2000, hayward_formation_2006, ansoldi_spherical_2008, chamseddine_nonsingular_2017, de_cesare_singularity_2020, babichev_regular_2020, cano_electromagnetic_2020} and \cite{ellis_ether_1973, bronnikov_scalar_1973, morris_wormholes_1988, visser_lorentzian_1996, li_morris-thorne_2020, sahoo_traversable_2020, maldacena_humanly_2020}. However, the discussion in \cite{carballo-rubio_opening_2020,carballo-rubio_geodesically_2020} proves that they are both theoretically motivated. Actually, it suggests more: if one's aim is to build phenomenological models that are agnostic e.g.~to the scale of regularisation, one would better not commit to one particular scenario but should rather prefer a unified treatment. Simpson and Visser's proposal fits perfectly in this line of reasoning; and it does so at no expense of simplicity --- a greatly attractive feature for the sake of phenomenology.

The capability of the metric \eqref{eq:SV} to properly describe realistic situations, however, is hindered by the lack of an important ingredient: rotation \cite{cardoso_testing_2019,malafarina_what_2020}. Not surprisingly, rotating regular black holes have been considered before, see e.g.~\cite{bambi_rotating_2013, neves_regular_2014, toshmatov_rotating_2014, dymnikova_regular_2015, ghosh_ergosphere_2020, kumar_shadow_2020}; similarly, rotating traversable wormholes have been proposed in \cite{teo_rotating_1998} and used for phenomenological modelling e.g.~in \cite{amir_shadow_2019, bueno_echoes_2018, gyulchev_shadow_2018, jusufi_gravitational_2018, tsukamoto_collisional_2015}. Predictably, the objects considered in the aforementioned references fall in two very distinct classes. The goal of this paper, therefore, is to construct a spinning generalisation of the Simpson--Visser metric \eqref{eq:SV} suitable for comparison with observations. To achieve it, we employ the Newman--Janis procedure, a method we introduce and describe below.

The paper is structured as follows. The Newman--Janis procedure is reviewed and applied in section \ref{sec:building}; the section terminates with our proposal for a spinning SV metric. Section \ref{sec:metric_analysis} is devoted to the global analysis of the ensuing spacetimes. Section \ref{sec:SEM} investigates the distribution of stress-energy that, assuming GR holds, produces our metric as a solution. Section \ref{sec:features} describes relevant features of the exterior geometry, namely ergoregion, photon ring and innmermost stable circular orbit. Finally, section \ref{sec:conclusions} reports our conclusions. We adopt units in which $c=G=1$, unless otherwise stated, and metric signature $(-, +, +,+)$.

\section{Building the metric}\label{sec:building}
    \subsection{The Newman--Janis procedure}
    
The Newman--Janis procedure (NJP) \cite{newman_note_1965} is a five-step method to build an axially symmetric spacetime starting from a spherically symmetric one. Its most remarkable application is the construction of the Kerr--Newman solution \cite{newman_metric_1965}; since then, several extensions to treat e.g. gauge fields or other charges have been studied and successfully employed --- see \cite{erbin_janis-newman_2017} for a summary. Notably, it has been used in \cite{bambi_rotating_2013} to generate spinning regular BHs. Despite its successes, the method encodes some puzzling arbitrariness   ---   on which we will elaborate further below   ---   and a true understanding of its working is still lacking. Some solid ground has been established in \cite{szekeres_explanation_2000}; see also \cite{rajan_complex_2015, rajan_cartesian_2017}. More recently, further insight has come from the study of scattering amplitudes \cite{arkani-hamed_kerr_2020, guevara_worldsheet_2020}.

For the reader's convenience, we briefly summarise here the NJP applied to a generic static, spherically symmetric seed metric
\begin{equation}
\dd s^2 = - f(r) \dd t^2 + \frac{\dd r^2}{f(r)} + h(r) [\dd \theta^2 + \sin^2 \theta \dd \phi^2];
\end{equation}
we will postpone specifying these results to (\ref{eq:SV}) until the next subsection.

As a step I, write the metric in outgoing (or, equivalently, ingoing) Eddington--Finkel\-stein coordinates $(u, r, \theta, \phi)$. Then (step II)
introduce a null tetrad $\{l^\mu, n^\mu, m^\mu, \overline{m}^\mu\}$ (the overline marks complex conjugation) satisfying $l^\mu n_\mu =-m^\mu \overline{m}_\mu = -1$ and $l^\mu m_\mu = n^\mu m_\mu=0$.
In terms of this tetrad, the metric can be written as
\begin{equation}
g^{\mu \nu} = -l^\mu n^\nu - l^\nu n^\mu + m^\mu \overline{m}^\nu + m^\nu \overline{m}^\mu.
\end{equation}

As a step III, define
\begin{equation}
r' = r+\iu a \cos\theta, \quad u' = u+\iu a\cos \theta, \quad \theta'=\theta, \quad \phi'=\phi,
\end{equation}
where $a$ is a real parameter to be identified, \textit{a posteriori}, with the spin; even though $r', u'$ are complex, these relations define a viable change of coordinates. The usual vector transformation law thus yields a transformed tetrad and, therefore, a transformed metric.

To obtain a new, axially symmetric metric, one needs (step IV) to replace the old functions $f,\ h$ with new $\tilde f,\ \tilde h$; the latter are required to be real, though of complex variable, and to coincide with the former when evaluated on the real axis. This replacement, in the standard NJP, is performed in a rather particular way but is, nonetheless, arbitrary.  For example, in the Schwarzschild geometry one has
\begin{equation}
f(r) = 1- \frac{2M}{r}, \quad h(r) = r^2;
\end{equation}
to derive the Kerr solution, $\tilde f,\  \tilde h$ are given by substituting (``complexifing'')
\begin{equation}
\frac{1}{r} \to \frac{1}{2} \bigg( \frac{1}{r'} + \frac{1}{\overline{r'}} \bigg), \quad r^2 \to r' \overline{r'}
\end{equation}
in $f$ and $h$  ---  all other prescriptions fail. We would like to stress that, generically, there is no real justification for the replacement above, except that the result of the NJP when applied to vacuum and electrovacuum solutions yields vacuum and electrovacuum solutions.

The complexification usually produces a metric with several non-diagonal components. To eliminate all of them, except for $g_{t\phi}$, one needs to perform an additional transformation to Boyer--Lindquist-like coordinates (step V). The desired change is of the form $\dd t' = \dd u - F(r)\dd r, \ \dd \phi' = \dd \phi - G(r) \dd r$, but is not always possible. Indeed, to integrate the above relations one needs $F, \ G$ to be independent from $\theta$: this is the case, for instance, in Schwarzschild and Reissner--Nordstr\"{o}m, but examples of the contrary exist.

To circumvent at once two problems of the NJP, namely the arbitrariness in the ``complexification'' and the possible lack of a transformation to Boyer--Lindquist-like coordinates, a modified version of the procedure (MNJP) was devised in \cite{azreg-ainou_generating_2014, azreg-ainou_static_2014}. In this framework, all the arbitrariness is encoded in a new function $\Psi$, which enters the resulting metric as an additional degree of freedom; the advantage with respect to the standard NJP is that a Boyer--Lindquist form is ensured (see also \cite{lima_junior_spinning_2020}) and one can thus set out to constrain $\Psi$ invoking physical arguments. In \cite{azreg-ainou_generating_2014, azreg-ainou_static_2014} this was done by demanding that the rotating metric be a solution of GR for an imperfect fluid, thence deriving a differential equation for $\Psi$.

In practice, however, given the extreme similarity of the metric (\ref{eq:SV}) with the Schwarz\-schild one, in our case all the potential issues that may prevent a successful implementation of the NJP are either easily worked around or outright absent.
    
\subsection{The rotating Simpson--Visser metric}
    
To apply the NJP to the metric (\ref{eq:SV}), i.e.\ to 
\begin{equation}
f(r) = 1-\frac{2M}{\sqrt{r^2 + \ell^2}}, \quad h(r) = r^2 + \ell^2,
\end{equation}
we need first of all to define (step I)
\begin{equation}
    \dd u := \dd t - \dd r_* := \dd t - \frac{\dd r}{f(r)}.
\end{equation}

The null tetrad that satisfies all requirements of step II is
\begin{equation}
    l^\mu = \delta^\mu _r, \quad n^\mu = \delta^\mu _u - \frac{f(r)}{2}\,\delta^\mu _r, \quad m^\mu = \frac{1}{\sqrt{2h(r)}} \left( \delta^\mu_\theta +\frac{\iu}{\sin\theta}\,\delta^\mu _\phi \right).
\end{equation}.

By changing coordinates to $u',\  r'$ (step III) we get
\begin{align}
    l'^\mu = l^\nu \pdv{x'^\mu}{x^\nu} = \delta^\mu_{r'}, \quad n'^\mu = \delta^\mu_{u'} -\frac{f(r)}{2}\,\delta^\mu_{r'}, \\
    m'^\mu = \frac{1}{\sqrt{2h(r)}} \left( \delta^\mu_{\theta'} -\iu a \sin\theta \left(\delta^\mu_{r'}-\delta^\mu_{u'}\right) +\frac{\iu}{\sin\theta}\,\delta^\mu _{\phi'} \right),
\end{align}
where $r$ is now meant as a scalar function (not a coordinate).

We now need to provide a prescription for the complexification (step IV) of $r$ yielding the new $\tilde f,\  \tilde h$. Since the components of the SV metric are derived from those of Schwarzschild by writing $\sqrt{r^2 + \ell^2}$ instead of $r$, there seems to be a natural choice: we can introduce a new coordinate $\varrho:= \sqrt{r^2 + \ell^2}$, and complexify it as would be appropriate for Schwarzschild's radial coordinate. Namely
\begin{equation}
    \varrho \to \varrho'= \varrho + \iu a \cos \theta,
\end{equation}
so that
\begin{equation}
 h (r) = \varrho^2 \to \tilde h(r') = \varrho' \overline{\varrho'} = r^2 + \ell^2 + a^2 \cos^2 \theta
\end{equation}
and
\begin{equation}
   f(r) = 1- \frac{2M}{\varrho} \to \tilde f(r') = 1- M\left(\frac{1}{\varrho'} + \frac{1}{\overline{\varrho'}}\right) =1- \frac{2M\sqrt{r^2 + \ell^2}}{r^2 + \ell^2 + a^2\cos^2\theta}.
\end{equation}

Now, the coordinate transformation of step V has the general form (see \cite{bambi_rotating_2013})
\begin{equation}
F = \frac{\tilde h (r, \theta) + a^2 \sin^2 \theta}{\tilde f(r, \theta)\tilde h(r, \theta) + a^2 \sin^2\theta}, \quad G = \frac{a}{\tilde f(r, \theta)\tilde h(r, \theta) + a^2 \sin^2\theta}.
\end{equation}
Specifying to our case:
\begin{align}
    F = \frac{r^2+\ell^2 + a^2}{r^2 + \ell^2 + a^2 -2M\sqrt{r^2 + \ell^2}},\quad
    G = \frac{a}{r^2 + \ell^2 + a^2 -2M\sqrt{r^2 + \ell^2}};
\end{align}
these expressions do not depend on $\theta$ and one can safely integrate them to get $t'(u,r), \ \phi'(u,r)$. (From now on we drop the primes.)

Thus, the metric obtained by applying the NJP to the SV seed does have a Boyer--Lindquist form. Note that this is obvious, in hindsight, since the functions $F, \ G$ above are the same that one would get starting from a Schwarzschild seed, provided one replaces the coordinate radius $r$ with $\sqrt{r^2 + \ell^2}$, and a Boyer--Lindquist form certainly exists in that case.

The metric ensuing from the application of the NJP with the choices above is our proposal for the rotating counterpart to the SV metric (\ref{eq:SV})
\begin{equation}
    \dd s^2 = -\left(1-\frac{2M \sqrt{r^2 + \ell^2}}{\Sigma} \right) \dd t^2 + \frac{\Sigma}{\Delta}\,\dd r^2 + \Sigma \dd \theta^2 - \frac{4M a \sin^2 \theta \sqrt{r^2 + \ell^2}}{\Sigma}\,\dd t \dd \phi + \frac{A \sin^2\theta}{\Sigma}\,\dd \phi^2 
    \label{eq:rotSV}
\end{equation}
with
\begin{gather*}
     \Sigma =  r^2 + \ell^2 + a^2 \cos^2 \theta, \qquad  \Delta = r^2 + \ell^2 + a^2 -2M \sqrt{r^2 + \ell^2}, \\
     A = (r^2 + \ell^2 + a^2)^2 - \Delta a^2 \sin^2 \theta.  
\end{gather*} 
It reduces to the SV metric when $a=0$ and to the Kerr metric when $\ell=0$. Formally, its components can be derived from those of the Kerr metric by replacing the Boyer--Lindquist radius $r$ with $\varrho = \sqrt{r^2 + \ell^2}$, but without changing $\dd r$; i.e.\ the metric \eqref{eq:rotSV} is \emph{not} related to Kerr by a change of coordinates.

The result (\ref{eq:rotSV}) obtained with the standard NJP is confirmed by the application of the MNJP, provided the arbitrary function $\Psi$ be fixed equal to $\Sigma$. This choice is coherent with requiring that the spinning metric coincides with Kerr when $\ell=0$, as it must.
    
The rest of the paper is devoted to characterising the metric (\ref{eq:rotSV}) and the spacetime it describes.    
    
\section{Metric analysis and spacetime structure} \label{sec:metric_analysis}

As in the non-spinning case, $r$ may take positive \emph{as well as negative} values. Our negative-$r$ region, however, should not be confused with the one deriving from analytically extending the Kerr spacetime beyond its ring singularity: indeed, the metric (\ref{eq:rotSV}) is symmetric under the reflection $r\to-r$ and the spacetime it describes is thus composed of two identical portions glued at $r=0$.

Some intuition can be gained by noting that the surface $r=0$ is an oblate spheroid of size (Boyer--Lindquist radius) $\ell$. When $\ell = 0$, the spheroid collapses to a ring at $\theta=\pi/2$ and the usual singularity of the Kerr geometry is recovered. When instead $\ell \neq 0$ the singularity is excised and $r=0$ is a regular surface of finite size, which observers may cross: the metric (\ref{eq:rotSV}) thus describes a wormhole with throat located at $r=0$. The nature of such throat (timelike, spacelike or null) depends on $\ell$ and $a$. Actually, $\abs{a}$ is the relevant parameter, thus, without loss of generality, we only consider $a>0$ here and throughout.

The values of $\ell$ and $a$ also determine whether the metric has coordinate singularities. When this is the case, the singularities are given by $\Delta = 0$ and located at
\begin{equation}
r_\pm = \left[\left(M \pm \sqrt{M^2-a^2}\right)^2 - \ell^2 \right]^{1/2}\,.
\end{equation}
By calling
\begin{equation}
    \varrho_\pm:= M \pm \sqrt{M^2 -a^2},
\end{equation}
we immediately see that $r_+$ is real only if $\ell\leq \varrho_+$ and, similarly, $r_-$ is real only if $\ell \leq \varrho_-$. Thus, depending on the values of the parameters, we may have two (if $a<M$ and $\ell< \varrho_-$), one (if $a<M$ and $\varrho_-<\ell < \varrho_+$) or no singularity at all (if $a<M$ and $\ell > \varrho_+$, or if $a>M$). The cases in which equalities hold are extremal or limiting versions of the above. As the analysis in the following subsection will prove, these coordinate singularities are horizons of the spacetime.

\subsection{Phase diagram\label{subsec:phase_diagram}}
\begin{figure}[tb]
    \centering
    \includegraphics[width=.8\textwidth]{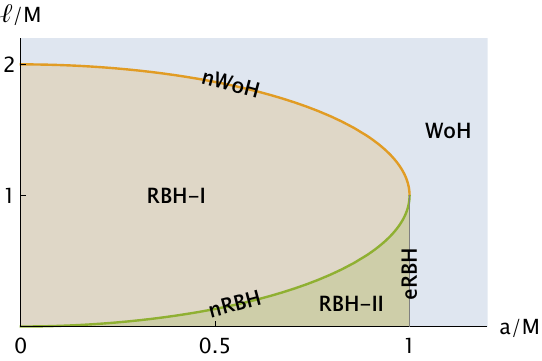}
    \caption{Parameter space and corresponding spacetime structure. Acronyms are spelled out in the text.}
    \label{fig:paramspace}
\end{figure}

For the sake of practicality, we summarise the spectrum of possible cases with the aid of a ``phase diagram'' in figure \ref{fig:paramspace}: each spacetime structure is associated with a region in (a constant-$M$ slice of) the parameter space under consideration. We defer a thorough discussion of each case to section \ref{subsec:penrose}, but lay out our terminology here:
\begin{description}
    \item[WoH] traversable wormhole;
    \item[nWoH] null WoH, i.e.\ one-way wormhole with null throat;
    \item[RBH-I] regular black hole with one horizon (in the $r>0$ side, plus its mirror image in the $r<0$ side);
    \item[RBH-II] regular black hole with an outer and an inner horizon (per side);
     \item[eRBH] extremal regular black hole (one extremal horizon per side);
    \item[nRBH] null RBH-I, i.e.\ a regular black hole with one horizon (per side) and a null throat.
\end{description}

A comment is in order, at this point. The parameter $\ell$ represents the spatial extent of the throat $r=0$. It seems difficult to envisage a single quantum gravity scenario capable of justifying all the values of $\ell$ that we consider. 

Indeed, if for instance one assumes gravitational collapse to proceed as predicted by GR until a certain threshold is met, beyond which quantum effects become dominant, one might deduce $\ell \sim L_\text{Planck}$ or $\ell \sim L_\text{Planck} \left(M/M_\text{Planck} \right)^{1/3}$. The two estimates correspond to different ways of identifying the threshold: in the first, quantum gravity becomes dominant when the radius of the collapsing object becomes of order the Planck length, in the second when the density is of order the Planck density \cite{rovelli_planck_2014, haggard_black_2015}. In either case, for astrophysically relevant masses, one would expect $\ell/M$ to be very small ($\sim 10^{-38} M_\odot /M$ in one case and $\sim 4\times10^{-26} \left(M_\odot /M\right)^{2/3}$ in the other) and RBH-II to be the only viable geometry, at least for $a \leq M$.

Values of $\ell/M = \order{1}$, instead, entail a macroscopic throat and most likely require additional ingredients. For instance, a regularisation at the Planck scale might be followed by a dynamical process, after which the structure settles down to the metric \eqref{eq:rotSV} \cite{malafarina_classical_2017}. If this is the case, there is no reason for $\ell$ to be linked to the scale of quantum gravity. Such process might preserve or destroy the horizon, so that the remnant object might correspondingly consist of a (regular) black hole or a ``naked'' wormhole. Note that $\ell/M =1$ is also the threshold above which traversable wormholes, with no horizons, can exist at spins $a<M$. 

We point out, in passing, that a similar ``phase diagram'' has recently been derived in \cite{brahma_testing_2020} by applying the NJP to a seed metric inspired by loop quantum gravity.

\subsection{Null rays and horizon structure}

To check that the singularities at $\Delta=0$ are coordinate artefacts, one can introduce ingoing null coordinates 
\begin{equation}
    \dd v:= \dd t + \frac{\varrho^2 + a^2}{\Delta}\,\dd r, \qquad \dd \psi:= \dd \phi + \frac{a}{\Delta}\,\dd r,
\end{equation}
and notice that the resulting metric is indeed regular at $r=r_\pm$ except perhaps when $v=\pm \infty$; equivalently, one could adopt outgoing null coordinates
\begin{equation}
    \dd u:= \dd t - \frac{\varrho^2 + a^2}{\Delta}\,\dd r, \qquad \dd \tilde \psi:= \dd \phi - \frac{a}{\Delta}\,\dd r,
\end{equation}
and confirm the same result except perhaps at $u=\pm \infty$. Either patch covers the region on which the other is not defined, thus proving that geodesics can be extended beyond $r=r_\pm$. The same deduction holds for $-r_\pm$.  

We further investigate the nature of the surfaces $r= \pm r_\pm$ by plotting the null rays $v= \cst,\  u= \cst$, in figure \ref{fig:nullrays}. We choose for simplicity $\theta=0$. The horizontal axes represents the Boyer--Lindquist radius $r$, while the time coordinate on the vertical axes $t_*^v$ is defined by
\begin{equation}
\dd t_*^v := \dd v - \dd r,
\end{equation}
so that 
\begin{align*}
v &= \cst \Rightarrow t_*^v = -r + \cst,\\
u &= \cst \Rightarrow t_*^v = -\int^r \left( 1-2\,\frac{\varrho^2 + a^2}{\Delta} \right) \dd r' + \cst.
\end{align*}
The peeling of outgoing rays shows that the surfaces $v = \cst$ and $r=\pm r_\pm$ are indeed horizons: $r_+$ and $-r_-$ are black-hole horizons, while $r_-$ and $-r_+$ are white-hole horizons. An analogous analysis adapted to outgoing rays, in which these appear as straight lines while ingoing rays present peeling, shows that the surfaces $\pm r_\pm$ and $u = \cst$ have the opposite nature with respect to their $v=\cst$ counterparts: $r_+$ is a white-hole horizon, $r_-$ a black-hole horizon, etc.

\afterpage{
\begin{figure}[htb]
   \centering
   \begin{subfigure}{.40\textwidth}
        \centering
        \includegraphics[width=\textwidth]{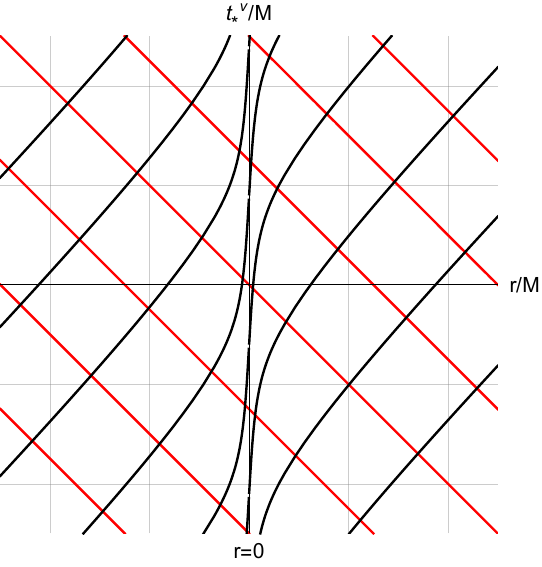}
        \caption{Traversable wormhole (WoH).}
    \end{subfigure} \qquad
    \begin{subfigure}{.40\textwidth}
        \centering
        \includegraphics[width=\textwidth]{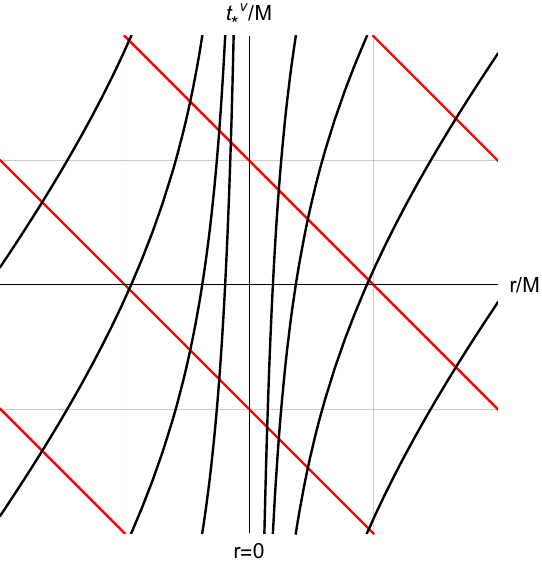}
        \caption{Null wormhole (nWoH).}
    \end{subfigure} \\
     \begin{subfigure}{.40\textwidth}
         \centering
        \includegraphics[width=\textwidth]{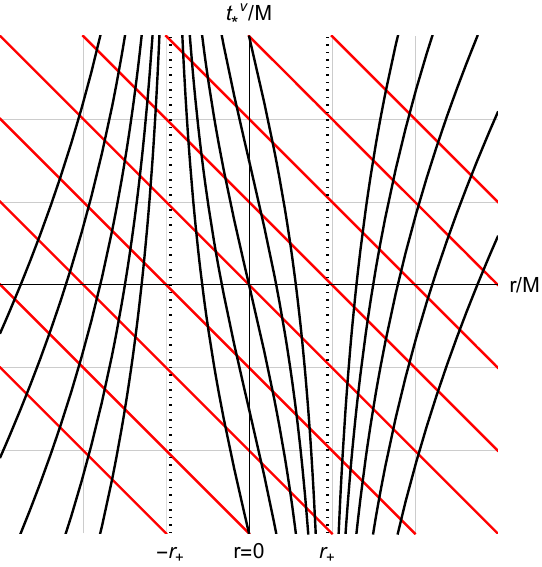}
        \caption{Regular black hole, one horizon (RBH-I).}
    \end{subfigure} \qquad
    \begin{subfigure}{.40\textwidth}
        \centering
        \includegraphics[width=\textwidth]{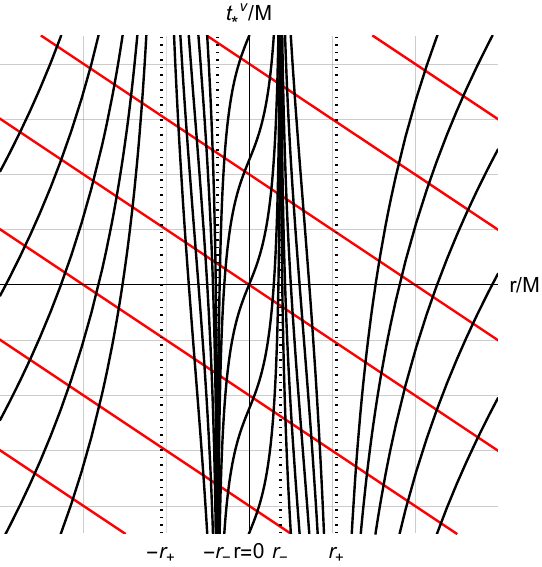}
        \caption{Regular black hole, two horizons (RBH-II).}
    \end{subfigure}\\
     \begin{subfigure}{.40\textwidth}
         \centering
        \includegraphics[width=\textwidth]{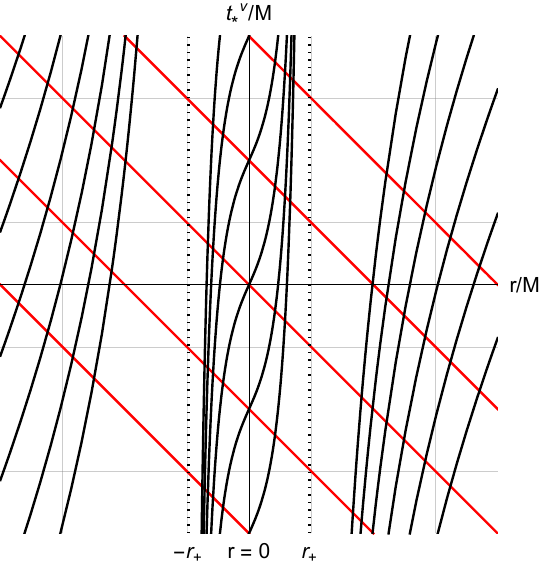}
        \caption{Extremal regular black hole, one horizon (eRBH).}
    \end{subfigure} \qquad
    \begin{subfigure}{.40\textwidth}
        \centering
        \includegraphics[width=\textwidth]{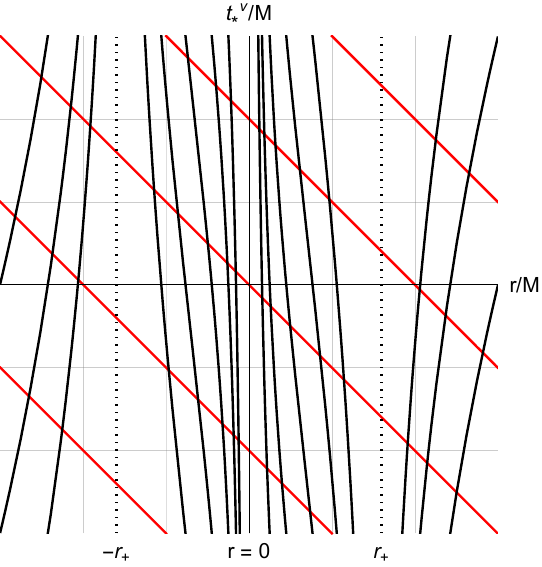}
        \caption{Null regular black hole, one horizon (nRBH).}
    \end{subfigure}
    \caption{Ingoing (red) and outgoing (black) null rays close to $\pm r_\pm$ for the different cases in the phase diagram. Particular values of $a$ and $\ell$ have been picked.}
    \label{fig:nullrays}
\end{figure} 
\clearpage}

\subsection{Carter--Penrose diagrams}\label{subsec:penrose}

The analytical extension of the metric (\ref{eq:rotSV}) across the horizons can be performed by standard methods (see e.g.\ \cite{chandrasekhar_mathematical_1983}), by changing to suitable Kruskal-like coordinates $U,\ V$ (a redefinition of $\phi$ is also required), defined in terms of $u,\ v$ by an exponential mapping involving the surface gravity of the horizon under consideration and compensating for the peeling of null rays off of it. 

The only practical difference between the textbook case of Kerr and our own lies in the functional relation between, e.g.\ the Boyer--Lindquist radius $r$ and the tortoise coordinate $r_*$. Such difference is inconsequential as far as analytic continuation is concerned; for instance, curves $UV = \cst$ correspond to curves $r=\cst$ in Kerr as well as in this case.

We construct Carter--Penrose diagrams for the maximal extension of the six cases identified in section \ref{subsec:phase_diagram} and report them in figures \ref{fig:penrose} and \ref{fig:penrose2}. A detailed description of each case follows.

\afterpage{
\begin{figure}[htb]
    \centering
\begin{subfigure}[t]{0.46\textwidth}
   \includegraphics[width=\textwidth]{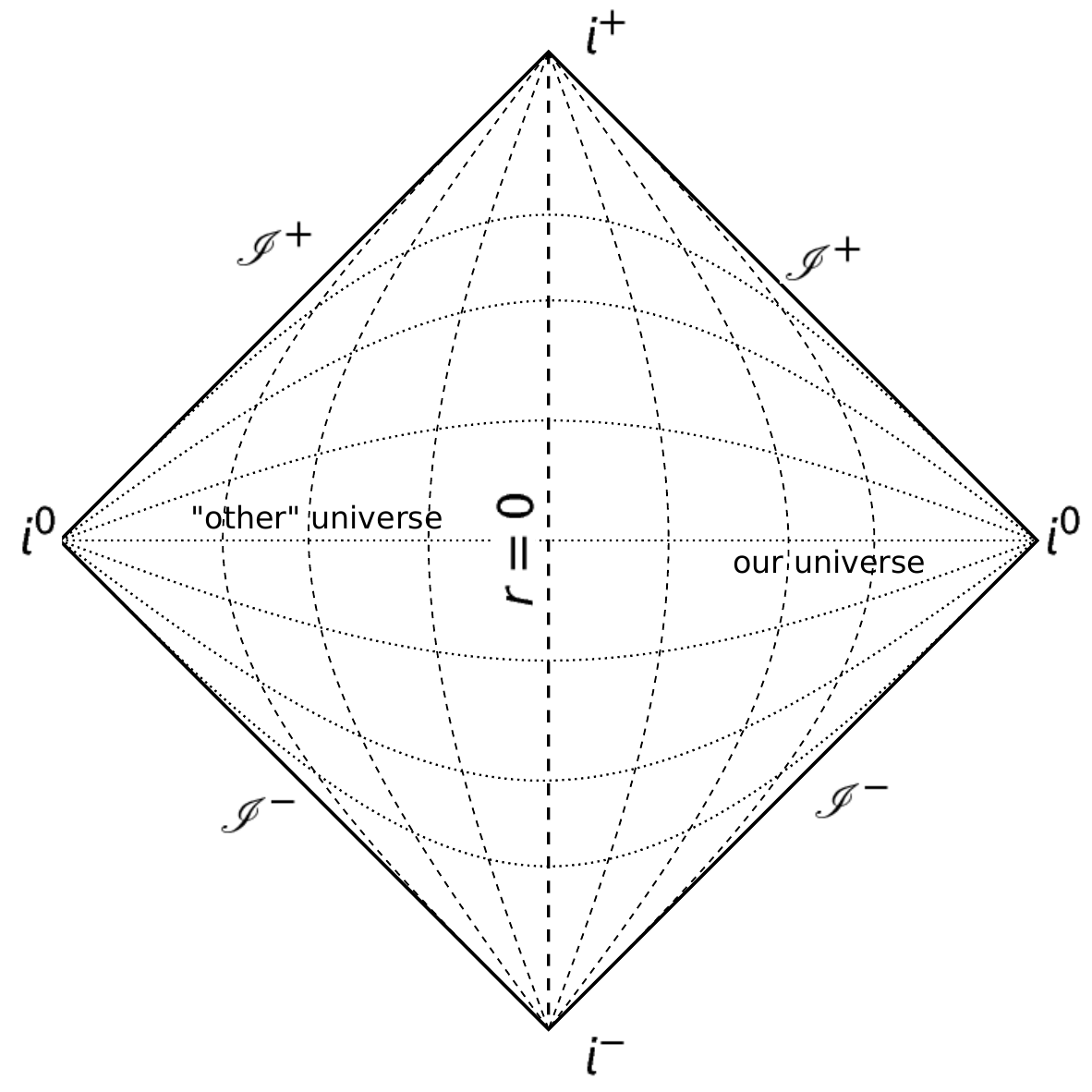}
   \caption{WoH, corresponding to $\ell > \varrho_+$ and $a<M$, or $a>M$. The throat $r=0$ is a timelike surface, traversable in both ways.}
\end{subfigure} \hfill
\begin{subfigure}[t]{0.46\textwidth}
  \includegraphics[width=\textwidth]{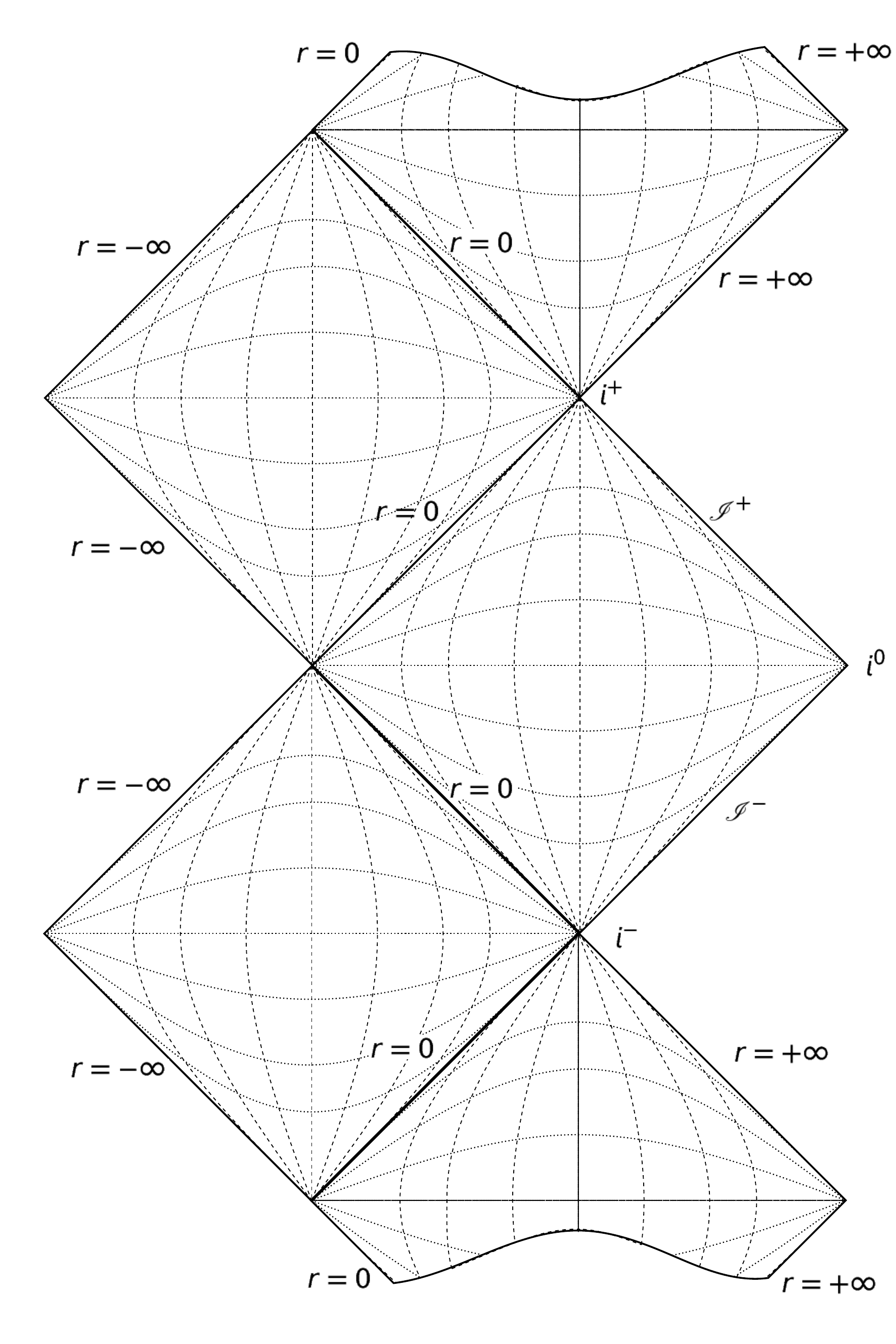}
   \caption{nWoH, corresponding to $\ell = \varrho_+$ (and $a<M$). The throat $r=0$ is a null surface and is an extremal event horizon.}
\end{subfigure}\\
\vspace{1em}
\begin{subfigure}[t]{0.46\textwidth}
   \includegraphics[width=\textwidth]{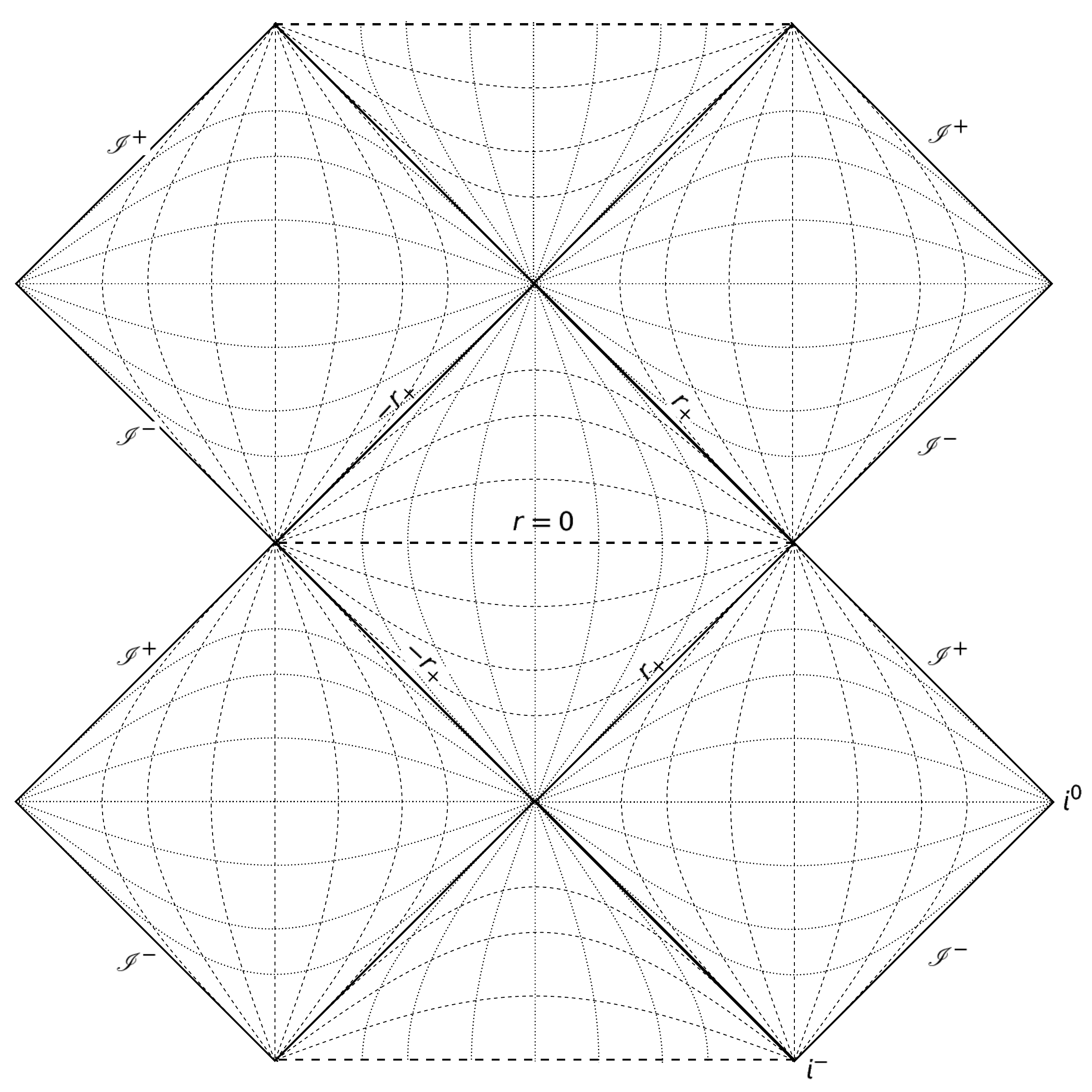}
   \caption{RBH-I, corresponding to $\varrho_- < \ell< \varrho_+$ (and $a<M$). The throat is spacelike and cloaked by an event horizon.}
\end{subfigure} \hfill
\begin{subfigure}[t]{.46\textwidth}
   \includegraphics[width=\textwidth]{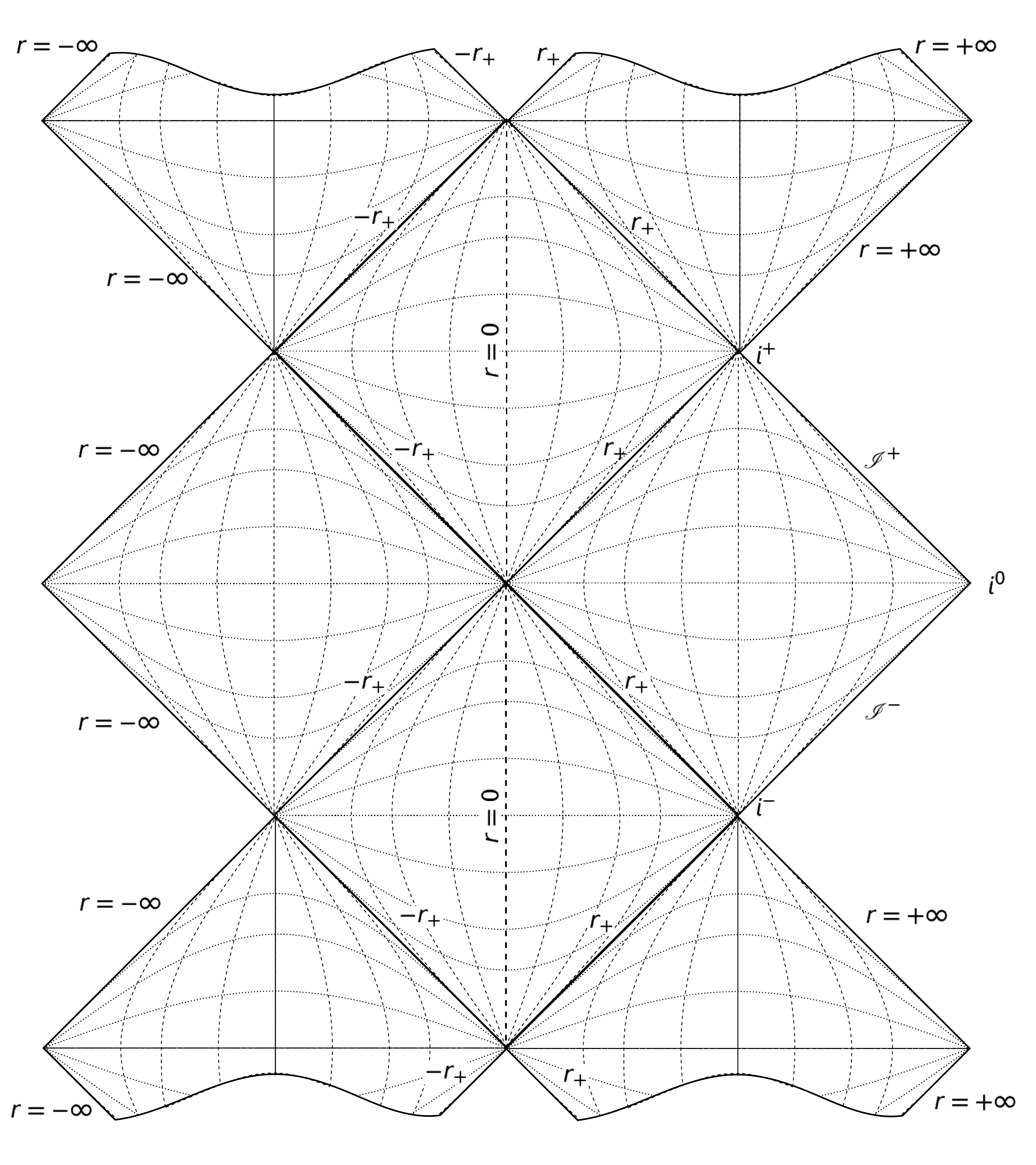}
   \caption{eRBH, corresponding to $a=M$ and $\ell < \varrho_- = \varrho_+$. The throat is timelike and the event horizon is extremal.}
\end{subfigure}
    \caption{Carter--Penrose diagrams for different spacetimes represented by the metric \eqref{eq:rotSV}. The lines $r=0$ correspond to the throat of the wormhole, which is a regular surface of finite area.}
    \label{fig:penrose}
\end{figure}

\begin{figure}[htb]
\centering
    \begin{subfigure}{\textwidth}
    \centering
    \includegraphics[width=.72\textwidth]{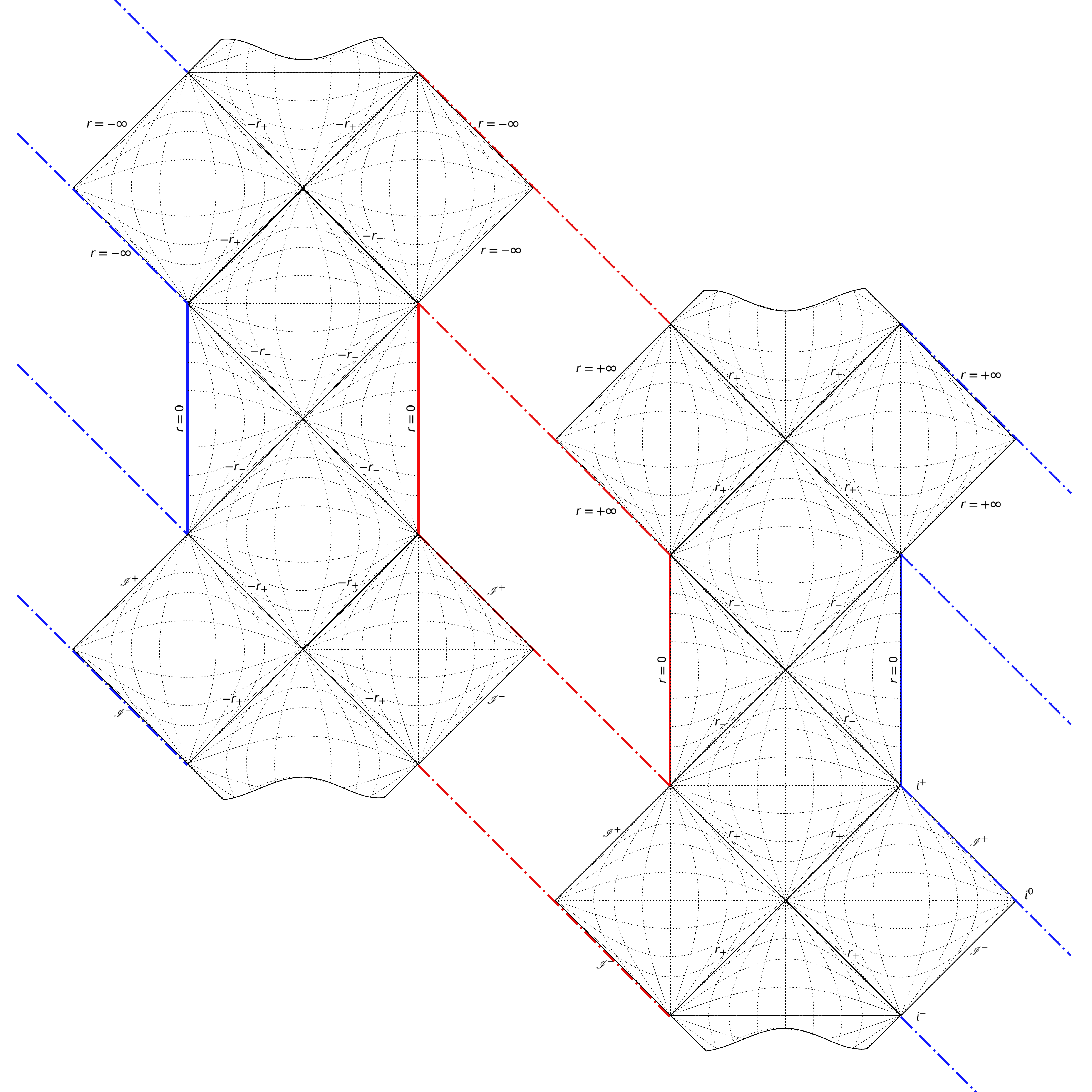}
    \caption{RBH-II, corresponding to $\ell < \varrho_-$ (and $a<M$). $r=0$ lines of the same colour, representing the wormhole throat, are identified.}
    \end{subfigure} \\
    \vspace{1em}
    \begin{subfigure}{\textwidth}
    \centering
    \includegraphics[width=.72\textwidth]{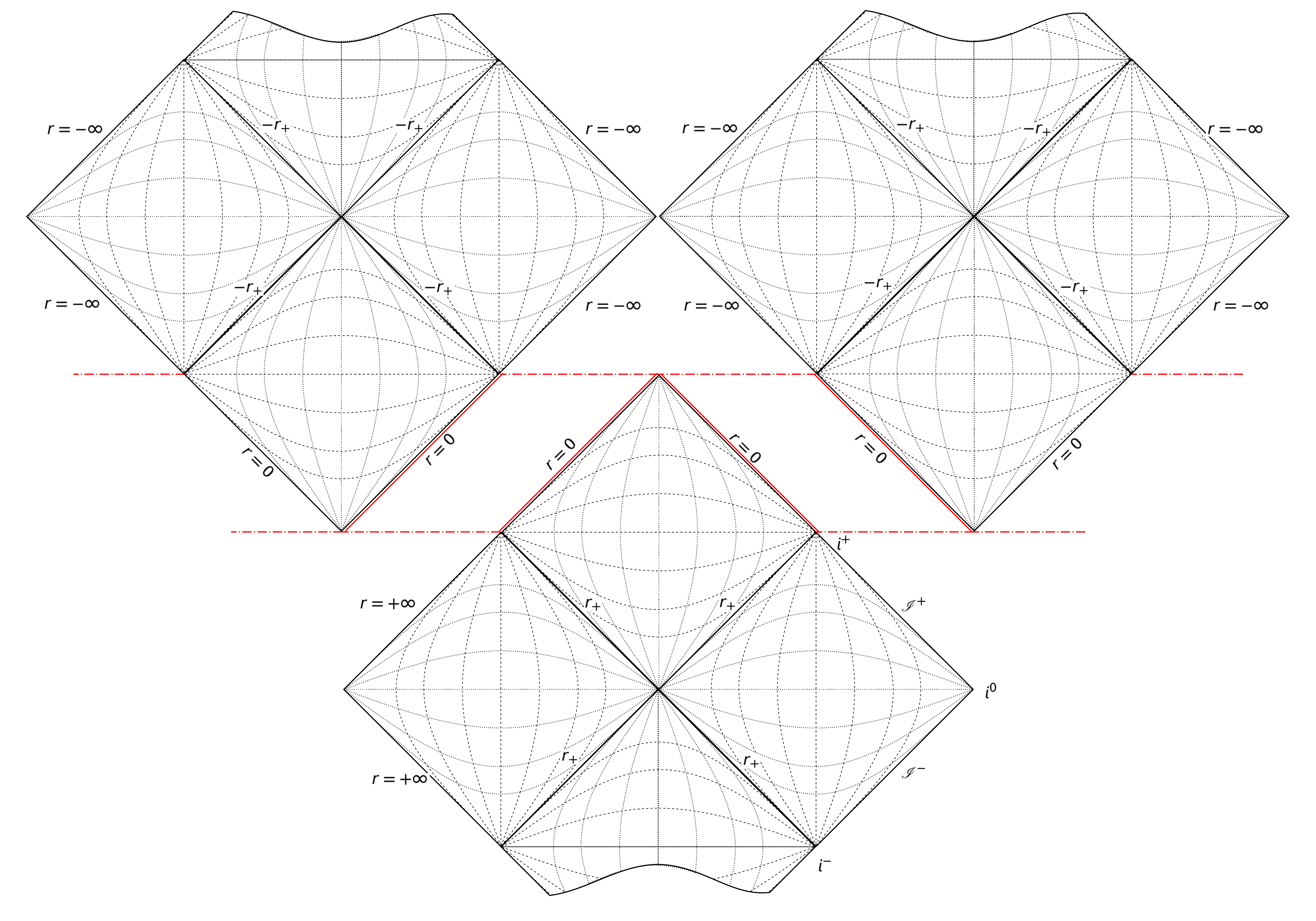}
    \caption{nRBH, corresponding to $\ell = \varrho_-$ (and $a<M$). Coloured $r=0$ lines are identified as indicated by dotted-dashed lines: they represent the wormhole throat, which in this case is also an extremal (inner) horizon.}
    \end{subfigure}
    \caption{Carter--Penrose diagrams (cont.). As in figure~\ref{fig:penrose}, $r=0$ corresponds to the regular, finite area throat.}
    \label{fig:penrose2}
\end{figure}
\clearpage}

\begin{description}
\item[WoH] is a traversable, two-way wormhole with a timelike throat. The Penrose diagram of this spacetime is the same as Minkowski's, provided one distinguishes the regions $r>0$  ---  ``our universe''  ---  and $r<0$  ---  the ``other universe''  ---  , since \textit{a priori} they are different. 

\item[nWoH] is a one-way wormhole with a null throat, which is an extremal event horizon. The analytically extended diagram continues indefinitely above and below the portion we show.

\item[RBH-I] is a spacetime containing an eternal black hole whose singularity is replaced by the (regular) throat of a spacelike wormhole. The diagram consists of infinitely many Schwarzschild-like blocks stacked one on top of the other and glued at the throats.

\item[RBH-II] is a regular black hole with two horizons per side. The Carter--Penrose diagram consists of two Kerr-like patches glued at the throats. The throats are timelike and can thus be traversed in both ways, yet the horizons are event horizons and an observer that crosses the throat twice can in no case return to the asymptotically flat region from which they left. Note that, from the point of view of an observer in ``our universe'', the inner horizon appears as a Cauchy horizon, as it does in the Kerr spacetime; here however, initial data cannot be specified in ``our universe'' only: Cauchy surfaces must be defined as disjoint unions of hypersurfaces that would be Cauchy in each of the Kerr-like patches. When one does so, one realises that $r_-$ (and $-r_-$) is not a Cauchy horizon and the spacetime is indeed globally hyperbolic.

\item[eRBH] is the extremal version of RBH-II, i.e.\ a regular black hole whose two horizons coincide. The surface gravity of the ensuing horizon is zero.

\item[nRBH] is the limiting case of RBH-I in which the throat becomes null; equivalently, it can be seen as the limiting case of RBH-II in which the throat coincides with the inner horizon. The diagram however looks markedly different from either case: it continues indefinitely above and below, as well as to the right and left; it consists of infinitely many fundamental blocks glued together at the throat. The null throat is an extremal (inner) horizon.
\end{description}

Note that the first three cases have already been analysed in \cite{simpson_black-bounce_2019}, while the last three are inherent to the rotating generalisation.

\subsection{Algebraic properties and surface gravity} \label{subsec:algebraic_properties}

The tetrad $\{l^\mu, n^\mu, m^\mu, \overline{m}^\mu\}$ resulting from the NJP coincides, in the case $\ell=0$, with the well-known Kinnersleay tetrad of the Kerr spacetime. 
The vector $l^\mu$, in particular, is geodetic and indeed tangent to the outgoing null rays analysed previously. $n^\mu$ is also geodetic but not affinely parametrised. 
One can use these vectors to build the tensor
\begin{equation}
    K^{\mu \nu}:= \Sigma \left(l^\mu n^\nu + l^\nu n^\mu\right) + \varrho^2 g^{\mu \nu}.
\label{eq:ktens}
\end{equation}
In Kerr, this is a Killing tensor; one can easily verify that it remains Killing even when $\ell \neq 0$. When contracted with a geodetic vector, it gives rise to a quantity that is conserved along the geodesic and can be used to define a generalisation of the so-called Carter constant of the Kerr spacetime --- see appendix \ref{append:notegeodesics} for details. As a consequence, the equations of motion for test particles are still separable.

One can further contract the tetrad with the Weyl tensor in order to determine the algebraic properties of this spacetime. We find
\begin{equation}
\Psi_0 = C_{\mu \nu \alpha \beta} l^\mu n^\nu l^\alpha n^\beta = 0
\end{equation}
(which agrees with \cite{szekeres_explanation_2000}, see proof of Theorem 2) and
\begin{equation}
\Psi_4 = C_{\mu \nu \alpha \beta} n^\mu \overline{m}^\nu n^\alpha \overline{m}^\beta =0,
\end{equation}
but 
\begin{align}
    \Psi_1 &= C_{\mu \nu \alpha \beta} l^\mu n^\nu l^\alpha m^\beta = \frac{a \ell^2 \sqrt{1-\chi^2}}{\sqrt{2} \iu \Sigma^2 (\varrho + \iu a \chi )},\\
    \Psi_2 &= C_{\mu \nu \alpha \beta} l^\mu m^\nu \overline{m}^\alpha n^\beta =
    -\frac{M}{\varrho  \Sigma^3}\left[\varrho^2\left(\varrho^2 -3 a^2 \chi^2\right) + \iu a \chi  \sqrt{\varrho^2-l^2} \left(3\varrho^2-a^2 \chi^2\right)\right]\0\\
    &\qquad\qquad + \frac{\ell^2}{6 \varrho^3 \Sigma^3} \left[2 a^2 \varrho ^2 \left(\chi ^2 (\varrho -4 M)-2 \varrho \right)-a^4 M \chi ^4+\varrho ^4 (9 M-2 \varrho )\right],\\
    \Psi_3 &= C_{\mu \nu \alpha \beta} l^\mu n^\nu \overline{m}^\alpha n^\beta = - \frac{a \ell^2 \sqrt{1-\chi^2} \Delta}{2\sqrt{2} \iu (\varrho - \iu a \chi) \Sigma^3},
\end{align}
where $\chi:=\cos\theta$. Thus this spacetime is not algebraically special.

The surface gravity that enters the analytic extension of section \ref{subsec:penrose} is
\begin{equation}
    \kappa_\pm:= \frac{1}{2} \dv{}{r} \eval{\bigg( \frac{\Delta}{\varrho^2 + a^2}\bigg)}_{r_\pm} = \kappa_\pm^\text{Kerr} \eval{\dv{\varrho}{r}}_{r_\pm} = \kappa_\pm^\text{Kerr} \sqrt{1-\frac{\ell^2}{\varrho_\pm^2}}.
\end{equation}
The expression above gives the so-called \textit{peeling} surface gravity. One can easily check that it agrees with the alternative, \textit{normal} definition 
\begin{equation}
\eval{\kappa^\pm_\text{normal}\, \Xi_\mu}_\mathcal{H^\pm}:= -\frac{1}{2} \nabla_\mu \eval{\bigg( \Xi^\nu \Xi_\nu \bigg)}_{\mathcal{H^\pm}},
\end{equation}
where $\mathcal{H^\pm}$ are the horizons and 
\begin{equation}
    \Xi^\mu \partial_\mu := \partial_t + \Omega\,\partial_\phi, \quad \Omega:= \frac{a}{\varrho^2 + a^2},
\end{equation}
is null and Killing at the horizons. Note in particular that the horizons are Killing and hence such coincidence of different definitions of surface gravity is expected~\cite{cropp_surface_2013}.

\section{Stress-energy-momentum and energy conditions} \label{sec:SEM}

In the context of GR, wormholes are typically associated with violations of the energy conditions \cite{visser_lorentzian_1996}. We thus characterise the distribution of ``effective'' matter entailed by the metric (\ref{eq:rotSV}) via the Einstein equations. We do so in two complementary ways: by focusing on particular geodesics (first a null congruence, then those of a timelike observer) and by diagonalising the Einstein tensor. The former method is more physical, in the sense that it sheds light on the energy density that actual observers would measure when orbiting these compact objects; the latter method is more systematic, in that it does not hinge on the  particular choice of observer. Note, in passing, our use of the modifier ``effective'': this analysis assumes Einstein's equations and might be severely reinterpreted in alternative theories of gravity (e.g. \cite{capozziello_energy_2014}). 

Further note that the throat is an extremum for the energy density, however defined. Indeed, for any $\varepsilon(r)$, 
\begin{equation}
\dv{\varepsilon}{r} = \dv{\varrho}{r} \dv{\varepsilon}{\varrho} = \frac{r}{\varrho} \dv{\varepsilon}{\varrho}
\end{equation}
and $r=0$ is automatically a zero when $\dd\varepsilon/\dd\varrho$ is finite. The sign of the second derivatives determines whether the extremum is a (local) minimum or maximum: denoting with a prime differentiation with respect to $\varrho$, we have 
\begin{equation}
\dv[2]{\varepsilon}{r} = \frac{r^2}{\varrho^2}\,\varepsilon'' +\frac{\ell^2}{\varrho^3}\,\varepsilon'.
\end{equation}

Solely for this section, we set $8\pi G \equiv 1$, so that $G_{\mu \nu}=T_{\mu \nu}$.

\subsection{Energy density for infalling observers}\label{subsec:ec_obs}
    
\paragraph{Null geodesics} Consider again the congruence $l^\mu$: being null and geodetic, these vectors are tangent to trajectories that fall towards the centre. Assuming GR holds, the contraction $G_{\mu \nu}l^\mu l^\nu$ is the energy density measured along these trajectories. A straightforward computation shows that
\begin{equation}
    G_{\mu \nu}l^\mu l^\nu = -\frac{2\ell^2}{\Sigma^2}.
\end{equation}

This quantity is negative, hence the null energy condition is violated. Thus exotic matter is encountered everywhere in the spacetime, in an amount that decreases as $1/r^4$ and is maximal at $r=0$. Note that the limit $\ell \to 0$ is zero at all radii and angles but $r=0,\  \theta=\pi/2$, where it is infinite; in fact
\begin{equation}
\eval{G_{\mu \nu}l^\mu l^\nu}_{r=0} = -\frac{2\ell^2}{\left(\ell^2 + a^2 \cos^2 \theta\right)^2}.
\end{equation}
Such behaviour is not surprising, since in the limit $\ell \to 0$ one recovers, at $r=0, \ \theta=\pi/2$, the standard ring singularity of the Kerr geometry. 

\paragraph{Timelike observer}\label{par:ECtobs} Consider now a timelike observer moving along a geodesic with tangent vector $u^\mu$. Because of the symmetries of this spacetime, the components $u_t=: - \mathcal{E}$ and $u_\phi =: \mathcal{L}$ are constants of motion, corresponding respectively to the observer's energy per unit mass and to the projection along the rotation axis of the observer's angular momentum per unit mass. Moreover, the existence of the Killing tensor (\ref{eq:ktens}) yields a third constant of motion, in terms of which $u_\theta$ can be expressed. The remaining component $u_r$ is fixed by the normalisation $u_\mu u^\mu = -1$.

We can restrict for definiteness to motion on the equatorial plane $\theta=\pi/2$, and compute again the double contraction with the Einstein tensor. We find:
\begin{equation}
    \varepsilon_u := \eval{G_{\mu \nu} u^\mu u^\nu}_{\theta=\pi/2} = -\frac{\ell^2}{\varrho^7} \big[ M(\mathcal{L}-a\mathcal{E})^2 -2\varrho(\mathcal{L}^2-a^2 \mathcal{E}^2) -\varrho^3(1-2 \mathcal{E}^2) \big].
    \label{eq:rhou}
\end{equation}
Thus  ---  assuming GR holds  ---  observers with, say, $\mathcal{L} = a \mathcal{E}$ measure a negative energy density at all radii when their energy is such that $\mathcal{E}^2 > 1/2$. The weak energy condition is therefore violated (and, consequently, the dominant energy condition too). Again, the limit $\ell \to 0$ yields zero except at the throat.

The sign of the second derivative of $\varepsilon_u$ at $r=0$ is the sign of $\left( x:= \mathcal{L}-a \mathcal{E} \right)$ 
\begin{equation}
\eval{\varepsilon_u'}_{r=0}= \eval{\ell^2\,\frac{7 M x^2-4 \varrho [6 a x \mathcal{E}+3 x^2+\varrho^2 \left(1-2 \mathcal{E}^2\right)]}{\varrho^8}}_{\varrho=\ell}.
\end{equation}
Thus, observers with $x=0$ and $\mathcal{E}^2>1/2$ measure at $r=0$ a (local) maximum of the energy density.

\subsection{Eigenvalue analysis}

To characterise the distribution of stress-energy in an observer-independent way, we diagonalise the Einstein tensor in mixed form\footnote{Note that $\tensor{G}{^\mu_\nu}$ is not symmetric, hence it is not guaranteed to be diagonalisable over the real numbers.} [eqs.~(\ref{Gcomponents}) in appendix \ref{append:curvatures}]. We find four distinct real eigenvectors, which in Boyer--Lindquist coordinates and up to multiplicative dimensionful constants are:
\begin{equation}
v_0^\mu = \{a + \varrho^2/a,0,0,1\}, \quad
v_1^\mu = \{0,1,0,0\}, \quad
v_2^\mu = \{0,0,1,0\}, \quad
v_3^\mu = \{a\sin^2\theta,0,0,1\}.
\end{equation}
We identify as minus the energy density $-\varepsilon$ the eigenvalue relative to the timelike eigenvector ($v_0^\mu$ when $\Delta >0$, $v_1^\mu$ otherwise); and as pressures $p_i$ ($i=1, 2, 3$) the other eigenvalues. We find 
\begin{subequations}\label{eq:rhop}\begin{align}
    \varepsilon &= \frac{\ell^2}{\Sigma^3} \bigg\{ \frac{\varrho \Sigma-2 a^2 M \chi^2}{\varrho} - 2 \Delta H \big( \Delta \big) \bigg\},\\
    p_1 & =\frac{\ell^2}{\Sigma^3} \bigg\{ \frac{ a^2 [2 M \chi^2-\varrho (\chi^2-2)]+\varrho^2 (\varrho-4 M)}{\varrho} - 2 \Delta H \big( \Delta \big) \bigg\},\\
    p_2 &= \frac{\ell^2}{\Sigma^3}\bigg\{ \frac{-M (2 a^2\chi^2\varrho^2+ \Sigma^2)+ \Sigma\varrho^3}{\varrho^3}\bigg\},\\
    p_3 &= -\frac{\ell^2}{\Sigma^3} \bigg\{\frac{a^2 \varrho^2 [4 M \chi^2+\varrho (\chi^2-2)]+a^4 M \chi^4+\varrho^4 (M-\varrho)}{\varrho^3}\bigg\},
\end{align}\end{subequations}
where $H (\cdot)$ is the Heaviside function. Note that the density and pressures so defined are continuous at the horizon, but their derivatives are generically not.

In the remainder of this section, we will analyse the null ($\varepsilon+p_i\geq 0$) and weak (null + $\varepsilon \geq 0$) energy conditions. Note however that
\begin{equation}
\varepsilon+p_1 = -\frac{2\abs{\Delta} \ell^2}{\Sigma^3} \leq 0
\end{equation}
and this suffices to prove that all energy conditions are violated (except possibly at the horizons). The other $\varepsilon+p_i$ have less wieldy expressions and we therefore not report them here.

Note that expressions (\ref{eq:rhop}) depend on the polar angle $\theta$ in a rather involved way but only through $\chi^2= \cos^2 \theta \in [0,1]$. We do not expect this dependence to induce dramatic features in the angular profiles of $\varepsilon$ and $p_i$; in particular, such dependence should be marked only at small radii and rapidly die out at spatial infinity. 

We confirm this intuition by studying $\varepsilon,\ p_i$ as functions of $\chi$, both analytically and graphically. The energy density, for instance, has an extremum at $\chi = 0$ (either a minimum or a maximum, depending on the values of the parameters); in addition, it may have at most two more extrema, symmetric with respect to $\chi=0$. Similar considerations apply to $\varepsilon+p_i$: $\chi=0$ is always an extremum and at most two other extrema, symmetric with respect to $\chi=0$, can exist. For $\varepsilon + p_1$, in particular, $\chi=0$ is the only extremum; at large radii it is a minimum  ---  but can become a maximum at smaller radii, depending on the parameters. 

\begin{figure}[tb]
   \centering
   \begin{subfigure}{.49\textwidth}
        \centering
        \includegraphics[width=\textwidth]{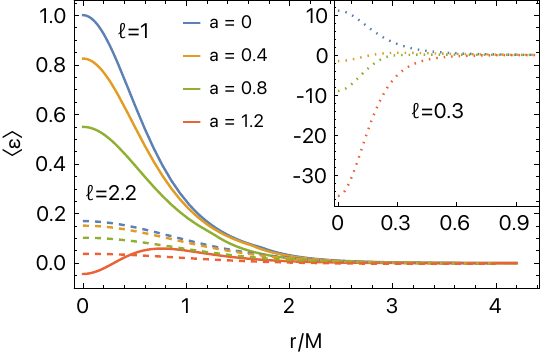}
    \end{subfigure} \hfill
    \begin{subfigure}{.49\textwidth}
        \centering
        \includegraphics[width=\textwidth]{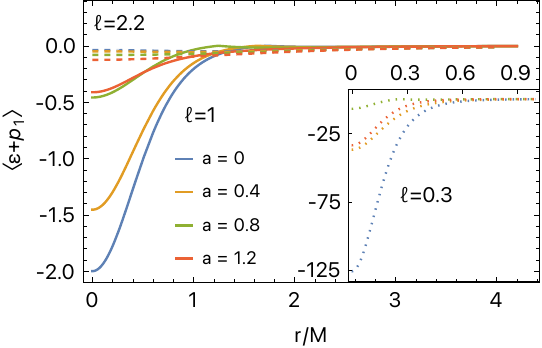}
    \end{subfigure}\\
    \vspace{.5cm}
     \begin{subfigure}{.49\textwidth}
         \centering
        \includegraphics[width=\textwidth]{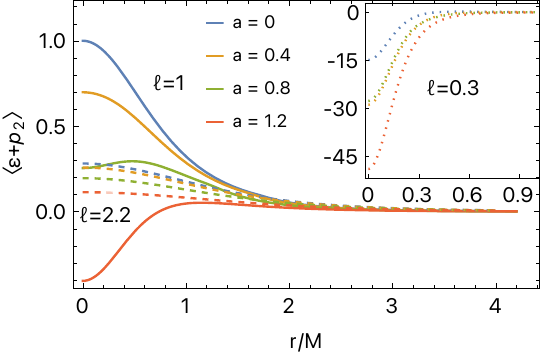}
    \end{subfigure} \hfill
    \begin{subfigure}{.49\textwidth}
        \centering
        \includegraphics[width=\textwidth]{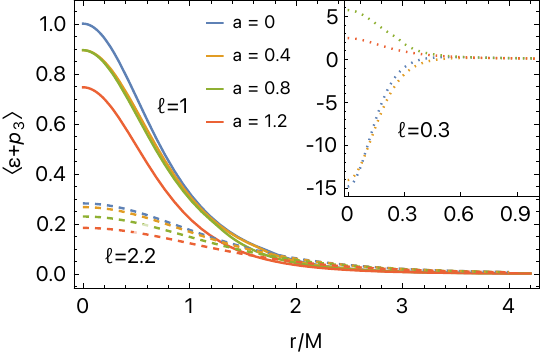}
    \end{subfigure}
    \caption{Energy conditions, averaged over the polar angle, as a function of $r$. Different colours represent different choices for the spin $a$, while different line styles stand for different values of $\ell$. Given the difference of scales, one particular value of $\ell$ is plotted in an inset.}
    \label{fig:ecrad}
\end{figure}

Hence, if we aim at characterising the radial distribution of stress and energy, a marginalisation over the angular variable is justified. Thus, for a generic quantity $X$, we resolve to consider
\begin{equation}
    \expval{X} := \frac{1}{2} \int_{-1}^{+1} \dd\chi\,X.
\end{equation}
In figure \ref{fig:ecrad} we plot $\expval{\varepsilon}$ and $\expval{\varepsilon+p_i}$ for selected values of $a$ and $\ell$. Notice that the stress-energy content of this spacetime is localised close to the origin: inspection of eq. (\ref{eq:rhop}) indeed confirms that energy density and pressures all scale as $1/r^4$. 

To quantify the amount of violation of the energy conditions in the whole spacetime, we adopt the strategy proposed in \cite{visser_traversable_2003,kar_quantifying_2004,nandi_volume_2004}. That is, we compute the so-called volume integral quantifier
\begin{equation}
    E:=\int \dd r \dd\theta \dd\phi\, \sqrt{\abs{g}}\,\varepsilon  , \quad E+P_i:=\int \dd r \dd\theta \dd\phi\, \sqrt{\abs{g}} \left(\varepsilon+p_i\right),
\end{equation}
where $g$ is the determinant of the four-dimensional metric.

\begin{figure}
   \centering
     \begin{subfigure}{.48\textwidth}
         \centering
        \includegraphics[width=\textwidth]{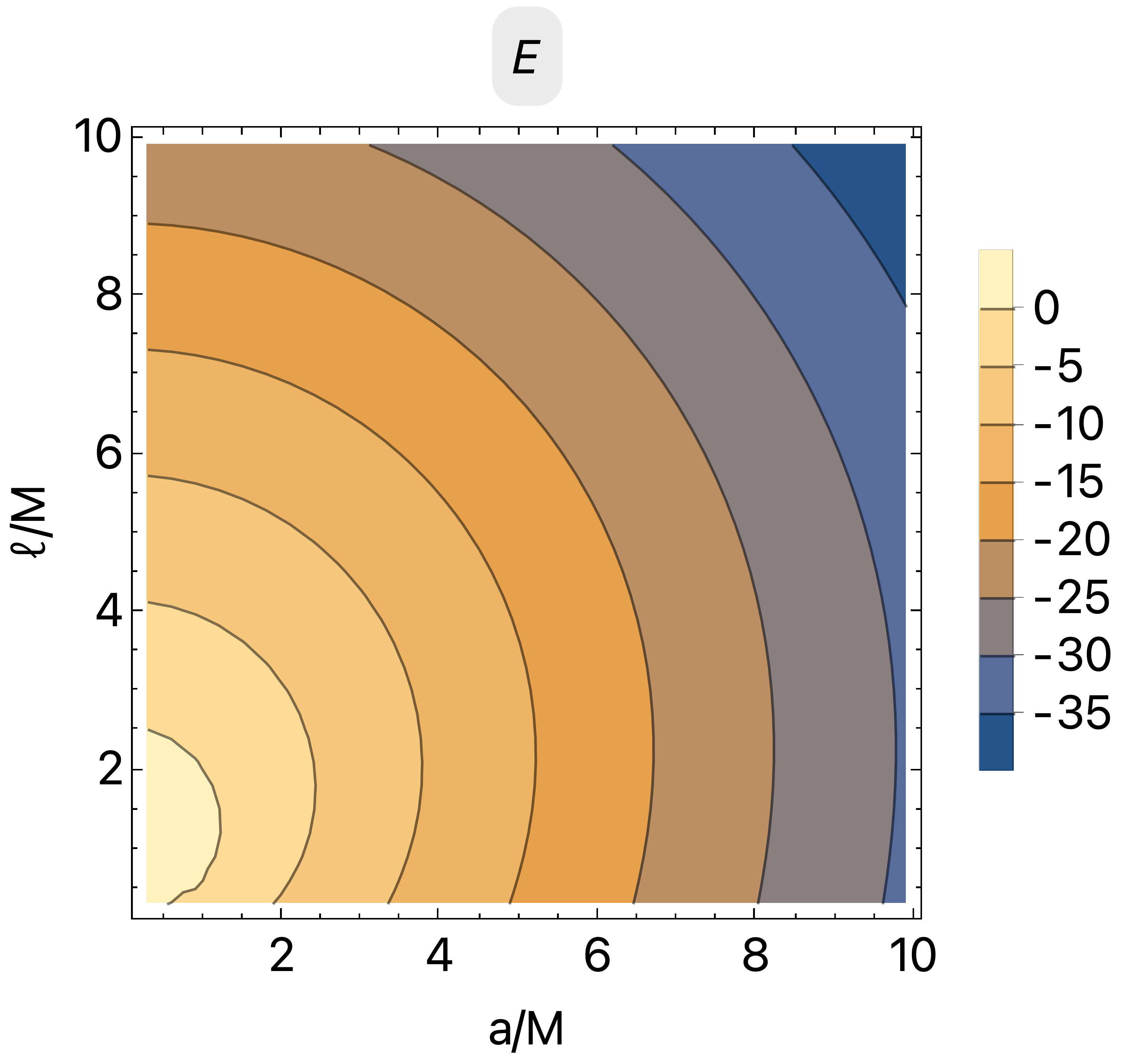}
        \label{fig:viqRho}
    \end{subfigure} \hfill
    \begin{subfigure}{.48\textwidth}
        \centering
        \includegraphics[width=\textwidth]{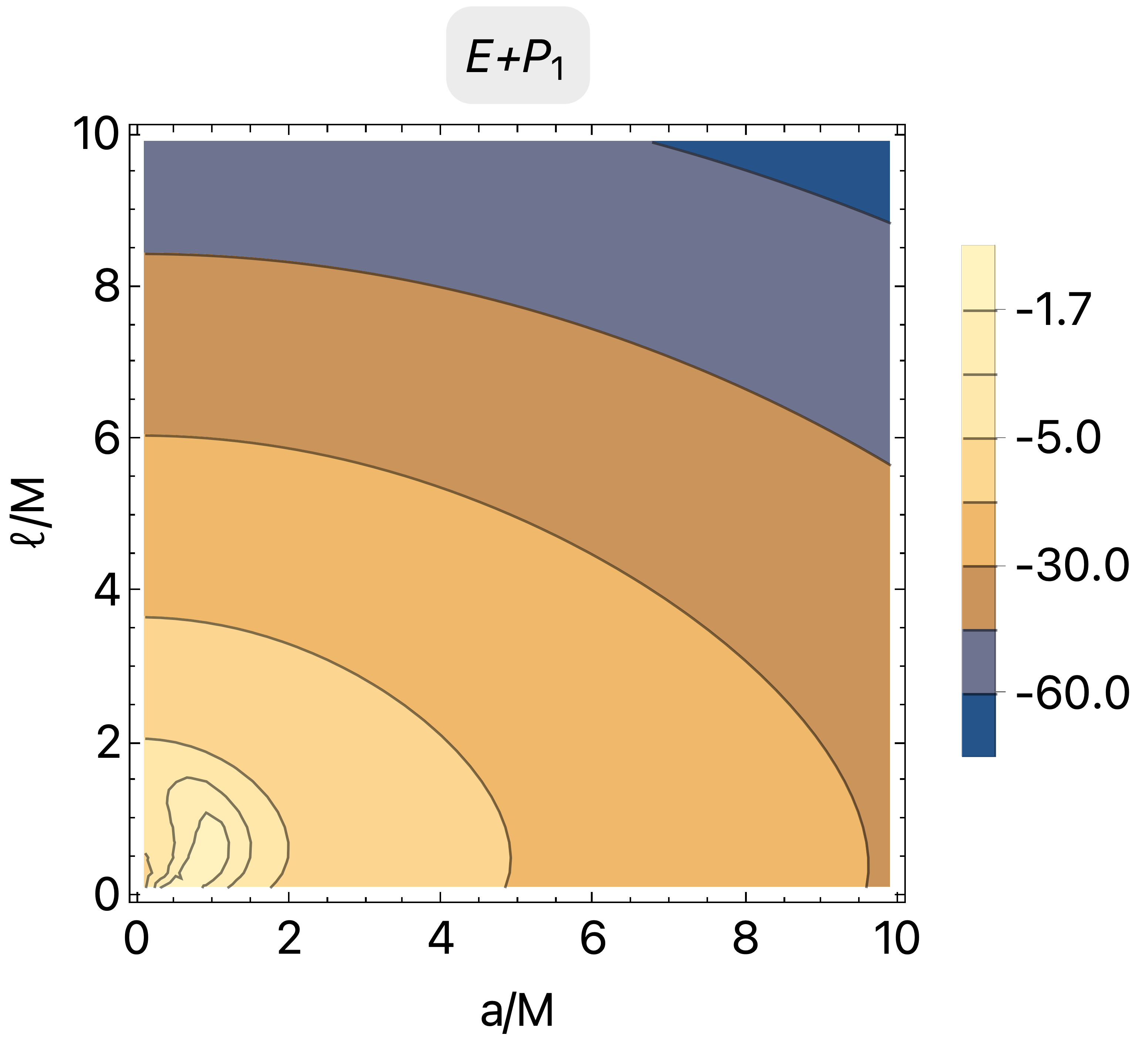}
        \label{fig:viqNEC1}
    \end{subfigure}\\
     \begin{subfigure}{.48\textwidth}
         \centering
        \includegraphics[width=\textwidth]{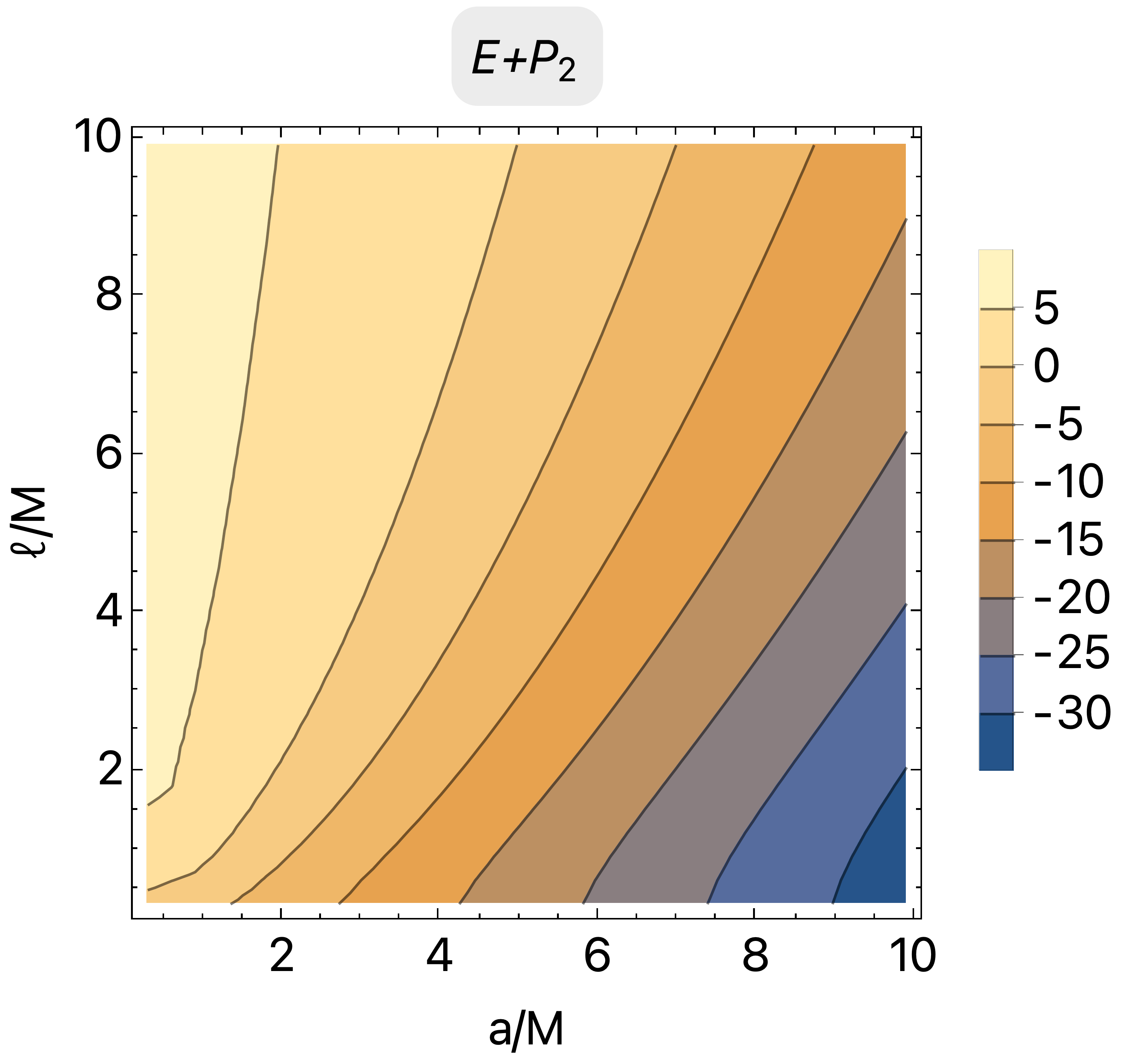}
        \label{fig:viqNEC2}
    \end{subfigure} \hfill
    \begin{subfigure}{.48\textwidth}
        \centering
        \includegraphics[width=\textwidth]{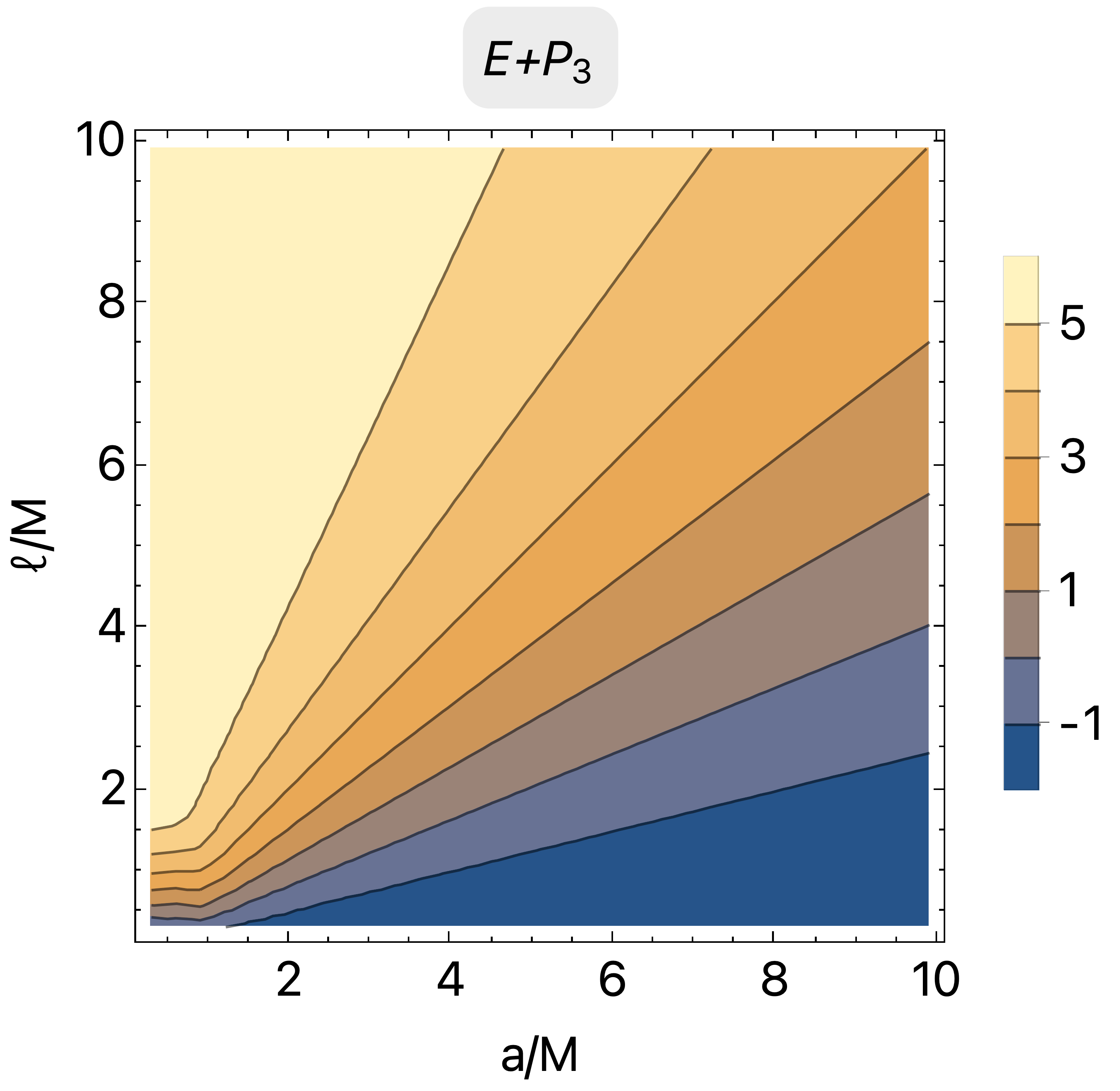}
        \label{fig:viqNEC3}
    \end{subfigure}
    \caption{Contour plots of the volume integral quantifier in the $(a,\ell)$-plane. Negative values of $E+P_i$ entail violations of the (averaged) null energy condition; negativity of $E$ or $E+P_i$ further entail violations of the (averaged) weak energy condition.}
    \label{fig:viq}
\end{figure}

We draw contour plots of these quantities for varying values of the parameters $a, \ \ell$ and report the results in figure \ref{fig:viq}. Note that the amount of ``effective'' matter varies substantially as the parameters vary and can be made very small by trimming them carefully.  

\section{Features of the exterior geometry} \label{sec:features}

The exterior of a Kerr black hole is rich in noticeable features, which largely determine the phenomenology of these objects. In this section, we focus on the exterior ($r\geq0$ and outside any horizon) of our geometry and study how switching on the parameter $\ell$ affects it. In particular, we describe the ergoregion and, schematically, the orbits, focusing on equatorial light ring and innermost stable circular orbit (ISCO).

\subsection{Ergoregion}
    
An ergoregion is a region inside of which no static observer can exist. Its boundary, the ergosurface, is the locus of points where $g_{tt}=0$. In our metric, the roots of this equation correspond to values of $\varrho$ given by

\begin{equation}
    \varrho_\text{erg}^\pm := M \pm \sqrt{M^2 - a^2\cos^2 \theta}.
    \label{eq:erg}
\end{equation}

Clearly, this expression coincides with what one finds in Kerr. Contrary to what is usually assumed in that case, however, here we do consider arbitrarily high spins. Therefore, the radicand in eq.~\eqref{eq:erg} is not always positive and the ergosurface has markedly distinct shapes in the $a>M$ and $a<M$ cases. An ergosurface exists when at least one of the quantities 
\begin{equation}
    r_\text{erg}^\pm = \sqrt{\left(\varrho_\text{erg}^\pm \right)^2 -\ell^2}
\end{equation}
is real. Note, incidentally, that $\min_{\theta \in [0, \pi]} \left(\varrho_\text{erg}^+ \right) = \varrho_+$ and $\max_{\theta \in [0, \pi]}\left(\varrho_\text{erg}^- \right) = \varrho_-$.

Hence, if $\ell \geq \varrho_\text{erg}^+$ there is no ergoregion. When this is the case, the object under consideration is a traversable wormhole (WoH case of section \ref{sec:metric_analysis}), since $\varrho_\text{erg}^+ \geq \varrho_+$; note that these wormholes may have arbitrary spin.

\begin{figure}[tb]
    \centering
    \begin{subfigure}[t]{.49\textwidth}
        \centering
        \includegraphics[width=\textwidth]{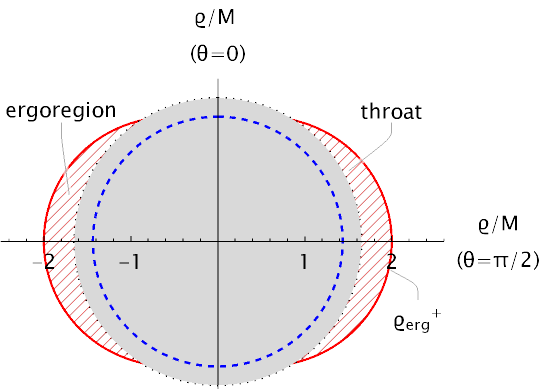}
        \caption{WoH with $a<M$. The ergoregion is limited to a crescent-shaped region about the equator. The dashed blue line represents the would-be outer horizon and is plotted for reference: the wormhole is traversable as long as the throat is ``larger than the horizon'', otherwise the object is a regular black hole (figure \ref{fig:ergoRBH}).}
        \label{fig:ergoWoHsub}
    \end{subfigure} \hfill
    \begin{subfigure}[t]{.49\textwidth}
        \centering
        \includegraphics[width=\textwidth]{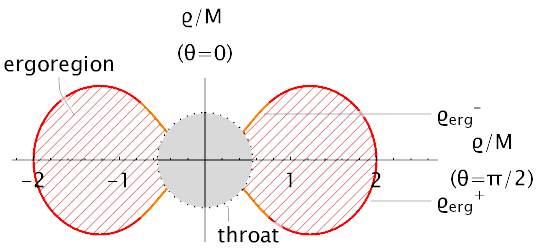}
        \caption{WoH with $a>M$. Note that the two branches $\varrho_\text{erg}^+$ and $\varrho_\text{erg}^-$, plotted in red and orange respectively, only exist for $\abs{\cos \theta} \leq a/M$ and join smoothly, thus producing an ergoregion of such peculiar shape.}
        \label{fig:ergoWoHsup}
    \end{subfigure}
    \caption{Traversable wormhole with ergoregion, corresponding to values of $\ell$ such that $\varrho_+ < \ell < \varrho_\text{erg}^+$. Each plot is a slice at fixed $\phi$, the angle with the vertical axis is $\theta$ and the distance from the centre is $\varrho$. The hatched region is the ergoregion. The black dotted line represents the throat $r=0$; the grey region is therefore \emph{excised} from the spacetime.}
    \label{fig:ergoWoH}
\end{figure}

If, on the contrary, $\ell < \varrho_\text{erg}^+$, an ergoregion is indeed present. This eventuality encompasses all cases of section \ref{sec:metric_analysis}, though with marked differences. 

Indeed, when the object is a traversable wormhole --- i.e.~if $a > M$, or $a<M$ and $\ell > \varrho_+$ ---, the throat intercepts the ergosurface at some angle $\theta \neq 0,\ \pi$; therefore, the ergoregion is limited to a region that is coaxial with the wormhole and whose longitudinal section is shaped as a crescent --- see figures  \ref{fig:ergoWoHsub} and \ref{fig:ergoWoHsup}. (This is a common feature of other rotating traversable wormholes, cf.~e.g.~\cite{teo_rotating_1998}). Note however that the throat is technically not an edge of the ergoregion, which in fact continues in the ``other universe'' as far as the mirrored ergosurface.

When instead the object is a regular black hole --- i.e.~if $a < M$ and $\ell \leq \varrho_+$ ---, the ergoregion extends all the way to the horizon and its external portion is thus tantamount to that of Kerr.

\begin{figure}[tb]
    \centering
    \begin{subfigure}[t]{.48\textwidth}
        \centering
        \includegraphics[width=\textwidth]{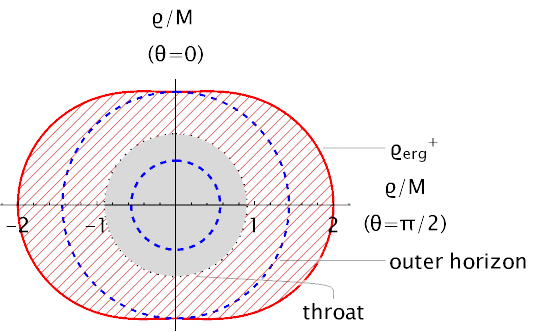}
        \caption{RBH-I (nWoH and nRBH are qualitatively similar). The ergoregion extends across the horizon and to the throat. The innermost dashed blue line represents the would-be inner horizon and is plotted for reference: the case in which the throat is ``smaller than the inner horizon'' is depicted in figure \ref{fig:ergoRBHII}.}
        \label{fig:ergoRBHI}
    \end{subfigure} \hfill
    \begin{subfigure}[t]{.48\textwidth}
        \centering
        \includegraphics[width=\textwidth]{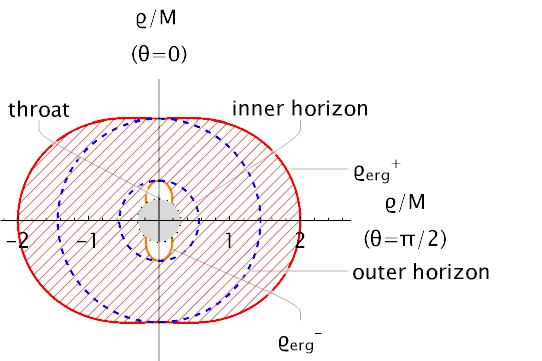}
        \caption{RBH-II (eRBH is qualitatively similar: the two horizons coincide and the inner and outer ergosurfaces touch at the poles). The ergoregion extends across two horizons, and to the throat or the inner ergosurface; within the white lobes $g_{tt}  < 0$, hence these regions do not belong to the ergoregion.}
        \label{fig:ergoRBHII}
    \end{subfigure}
    \caption{Ergoregion in regular black holes, i.e.\ when $\ell \leq \varrho_+$. As in figure \ref{fig:ergoWoH}, each plot is a slice at fixed $\phi$, the angle with the vertical axis is $\theta$ and the distance from the centre is $\varrho$. The hatched region is the ergoregion. The black dotted line represents the throat $r=0$; the grey region is therefore \emph{excluded} from the spacetime.}
    \label{fig:ergoRBH}
\end{figure}

For completeness, however, we describe the structure of the ergoregion inside the horizon, too. If $\varrho_- \leq \ell < \varrho_+$, viz.\ in the cases nWoH, RBH-I and nRBH, the ergoregion stretches as far as the throat --- see figure \ref{fig:ergoRBHI}. If instead $\ell < \varrho_-$, that is in the RBH-II and eRBH cases, the ergoregion has an inner ergosurface; this surface is intercepted by the throat at some angle $\theta$: thus the only portions of the spacetime, close to the throat, that do not belong to the ergoregion are lobes enclosing the poles --- see figure \ref{fig:ergoRBHII}. Note that, as before, the throat is not an edge of the ergoregion, in the sense that $g_{tt}$ does not change sign upon crossing it.

\subsection{Notable orbits}\label{subsec:orbits}
    
The study of orbits can proceed as for the Kerr geometry. More detail can be found in appendix~\ref{append:notegeodesics}; here we focus on the most relevant case of equatorial circular motion.

Indeed, the $t$- and $\phi$-motion are readily integrated by exploiting the conservation of the test particle's energy $\mathcal{E}$ and angular momentum $\mathcal{L}$ (per unit mass) along the rotation axis. Thus, setting $\theta=\pi/2$, the problem is effectively one-dimensional and governed by
\begin{equation}
    \varrho^2 \Dot{r} = \pm \sqrt{\mathcal{R}}
\end{equation}
where the dot marks differentiation with respect to the affine parameter along the geodesic and $\mathcal{R}$ is the same potential one finds for a Kerr spacetime mapped by the Boyer--Lindquist radius $\varrho$:
\begin{equation}
    \mathcal{R} = [\mathcal{E} (\varrho^2 + a^2) -a\mathcal{L}]^2 - \Delta[\mu^2 \varrho^2 + (\mathcal{L} -a \mathcal{E})^2],
\end{equation}
with $\mu^2=0, \ +1$ for null and timelike orbits, respectively (the expression for generic $\theta$ can be found in appendix \ref{append:notegeodesics}).

Circular orbits satisfy simultaneously
\begin{equation}
    \mathcal{R} = 0 \quad \text{and} \quad \dv{\mathcal{R}}{r} = 0.
    \label{eq:circular}
\end{equation}
Note that
\begin{equation}
    \dv{\mathcal{R}}{r} = \dv{\varrho}{r}\dv{\mathcal{R}}{\varrho} = \frac{r}{\varrho}\dv{\mathcal{R}}{\varrho},
\end{equation}
hence Kerr's circular orbits readily correspond to circular orbits of our spacetime. Since $r=0$ does not generically satisfy the above relations, the mapping between Kerr equatorial circular geodesics and our own is onto.

Thus, in particular, to any pair $\mathcal{E}_c,\  \mathcal{L}_c$ there corresponds a solution $\varrho_c$ of (\ref{eq:circular}) as long as 
\begin{equation}
    \varrho_c^2-3M \varrho_c \pm 2a \sqrt{M \varrho_c} \geq 0,
\end{equation}
where the plus (minus) sign refers to prograde (retrograde) orbits. When the equality holds, the equation has formally three real solutions for $a<M$ and only one for $a>M$. In the former case, however, the smallest of such roots lies inside the horizon and, therefore, does not correspond to any orbit; the other two correspond to the familiar unstable circular photon orbits $\varrho_\text{ph}$, one prograde and one retrograde, of the Kerr spacetime. In the latter case, the corresponding orbit connects smoothly to the retrograde branch of the $a<M$ case. These orbits of the Kerr spacetime translate into orbits of our spacetime, located at
\begin{equation}
    r_\text{ph} = \sqrt{\varrho_\text{ph}^2 - \ell^2}.
\end{equation}
Notice that for $\ell>3M$ we find no prograde photon circular orbit, at any spin; these wormholes however do have a retrograde circular orbit, if they spin fast enough.

Timelike circular orbits are stable as long as
\begin{equation}
    \varrho_c^2-6M \varrho_c \pm 8a \sqrt{M \varrho_c} -3a^2 \geq 0. 
\end{equation}
Once again, the equality gives rise to two branches of solutions for $a<M$, the prograde and retrograde branches, one of which (the retrograde) continues to the $a>M$ region. In the Kerr spacetime, these solutions represent the ISCO\@. In our spacetime, they are located at
\begin{equation}
r\indices{_{\text{ISCO}}} = \sqrt{\varrho\indices*{_{\text{ISCO}}^2} - \ell^2}.
\end{equation}
Note that, for $\ell > 6M$, these wormholes do not present prograde ISCO\@. They may have a retrograde ISCO, if they spin fast enough.

\section{Conclusions and outlook} \label{sec:conclusions}

In this paper, we have constructed a rotating generalisation of the Simpson--Visser metric, applying the Newman--Janis procedure. Depending on the values of $a$ and $\ell$, it may represent a traversable wormhole, a regular black hole with one or two horizons, or three limiting cases of the above. 
The global properties of the ensuing spacetime have been discussed at length. We further characterised our metric by describing the violations of the energy conditions and found that the ``exotic'' matter is localised in the vicinity of the throat. Finally, we investigated some relevant features of the exterior geometry: an ergoregion exists when $\ell < 2M$, whatever the value of the spin; a (retrograde) circular photon orbit exists for $\ell < 3M $ and a (retrograde) ISCO for $\ell<6M$, again independently on the spin.

The metric (\ref{eq:rotSV}) thus describes a family of reasonable Kerr black hole mimickers, suitable for serious phenomenological inquiry. For instance, one may wonder whether the presence of an ergoregion affects its stability. Or, again, whether $\ell \neq 0$ leads to observational consequences e.g.\ on the electromagnetic shadow. Investigation on these matters is under way.

\acknowledgments
We thank Matt Visser for a careful reading of the manuscript and for his precious comments.
The authors acknowledge funding from the Italian Ministry of Education and Scientific Research (MIUR) under the grant PRIN MIUR 2017-MB8AEZ.

\appendix

\section{Curvatures}\label{append:curvatures}

Curvature tensors and scalars are readily computed (we employed the \texttt{xAct} bundle \cite{martin-garcia_xact_2002, martin-garcia_xtensor_2002, yllanes_xcoba_2005} for \textsc{Mathematica}). Their expressions are not particularly illuminating, but we report some of them in order to make a few relevant comments.

The Ricci scalar reads ($\chi:= \cos \theta$):
\begin{equation}
    R = \frac{2 \ell^2}{\varrho^3 \Sigma^3} \bigg(a^2 \varrho^2 \big[2 M \chi^2+\varrho  (\chi^2-2) \big]+a^4 M \chi^4+\varrho^4 (3 M-\varrho )\bigg).
\end{equation}
For $a=0$ this expression coincides with Simpson and Visser's \cite{simpson_black-bounce_2019}; the limit $\ell \to 0$ is well defined and vanishing everywhere, except at $r=0$ and $\theta = \pi/2$. Recall that $\Sigma = r^2 + \ell^2 + a^2 \cos^2\theta$, hence the throat $r=0$ is not singular (not even at $\theta = \pi/2$) as long as $\ell \neq 0$. Note in particular that $r=0$ is an extremum point for the Ricci scalar, i.e.\ $\eval{\dv{R}{r}}_{r=0}=0$ (a minimum, at least for small $a, \ell$). 

The Kretschmann scalar reads:
\begin{align}
  R^{\mu \nu \lambda \sigma}R_{\mu \nu \lambda \sigma} &= \frac{48 M^2}{\Sigma^6}\,\mathcal{K}_0
+\frac{16 \ell^2 M}{\Sigma^6}\,\mathcal{K}_1 + \frac{4 \ell^4}{\varrho^6\Sigma^6}\,\mathcal{K}_2,
\end{align}
with
\begin{subequations}\begin{align}
  \mathcal{K}_0 &= (\varrho^2 - a^2 \chi^2) \left[\left(\varrho^2 + a^2 \chi^2\right)^2 - 16 a^2 \varrho^2 \chi^2 \right],\\
  \mathcal{K}_1 &= \varrho^3 \left[4 a^2+\varrho  (2 \varrho -9 M)\right]-2 a^2 \varrho  \chi^2 \left[6 a^2+\varrho  (4 \varrho -31 M)\right]+a^4 \chi^4 (6 \varrho -41 M),\\
  \mathcal{K}_2 &= \varrho^6 \left[2 a^2 \varrho  (2 \varrho -11 M)+4 a^4+\varrho^2 \left(33 M^2-16 M \varrho +3 \varrho^2\right)\right]\0\\
  &\phantom{=}+2 a^2 \varrho^5 \chi^2 \left[2 a^2 (5M-\varrho)-26 M^2 \varrho +5 M \varrho^2+\varrho^3\right] +2 a^6 M \varrho^2 \chi^6 (6 M-\varrho)\0\\
  &\phantom{=}+a^4 \varrho^3 \chi^4 \left[2 a^2 M+\varrho  \left(34 M^2-16 M \varrho +3 \varrho^2\right)\right] +a^8 M^2 \chi^8,
\end{align}\end{subequations}
again, for $r=0$  ---  i.e.\ $\varrho= \ell$  ---  this expression is finite as long as $\ell \neq 0$.

The components of the Einstein tensor, in mixed components, are: 
\begin{subequations}\label{Gcomponents}\begin{align}
    \tensor{G}{^t_t} &= \frac{\ell^2}{\varrho^3 \Sigma^4}\Big[ a^4 \varrho^2 \chi^2 [ 2M (3-2 \chi^2)-\varrho (\chi^2-2) ] + a^2 \varrho^4 [M (\chi^2-3)+2 \varrho ]\0\\
    &\phantom{=}- a^6 M \chi^4 (\chi^2-1)+\varrho^6 (\varrho -4 M)\Big],\\
    \tensor{G}{^\phi_t} &= \frac{a \ell^2 M (6 a^2 \varrho^2 \chi^2+a^4 \chi^4-3 \varrho^4)}{\varrho^3 \Sigma^4},\\
    \tensor{G}{^t_\phi} &= \tensor{G}{^\phi_t} \left(a^2 +\varrho^2\right)\left(\chi^2-1\right),\\
    \tensor{G}{^r_r} &= -\ell^2\frac{\varrho^3 - 2 a^2 M \chi^2+a^2 \varrho  \chi^2}{\varrho  \Sigma^3},\\
    \tensor{G}{^\theta_\theta} &= \ell^2 \frac{\varrho^5 - M (4 a^2 \varrho^2 \chi^2+a^4 \chi^4+\varrho^4)+a^2 \varrho^3 \chi^2}{\varrho^3\Sigma^3},\\
    \tensor{G}{^\phi_\phi} &= \ell^2 \frac{a^4 \varrho^2 \chi^2 [M (\chi^2-6)-\varrho  (\chi^2-2)]+a^2 \varrho^4 [M (3-8 \chi^2)+2 \varrho ] - a^6M \chi^4+\varrho^6 (\varrho -M)}{\varrho^3 \Sigma^4}.
\end{align}\end{subequations}

\section{A note on geodesics}\label{append:notegeodesics}

Geodesics can be described by means of the Hamilton--Jacobi method \cite{chandrasekhar_mathematical_1983, frolov_black_1998}, whereby the equations of motion descend from
\begin{equation}
    \pdv{S}{\tau} = -\frac{1}{2}\,g^{\mu \nu}\pdv{S}{x^\mu} \pdv{S}{x^\nu},
    \label{eq:hamilton-jacobi}
\end{equation}
with $S$ the action and $\tau$ an affine parameter along the geodesic. Separability, foretold in section~\ref{subsec:algebraic_properties}, motivates the ansatz
\begin{equation}
    S = \frac{1}{2}\,\mu^2 \tau - \mathcal{E}t + \mathcal{L} \phi + S_r(r) + S_\theta(\theta);
    \label{eq:action}
\end{equation}
here $\mu^2, \ \mathcal{E}$ and $\mathcal{L}$ are arbitrary constants: $\mu^2=0$ for null geodesics and $+1$ for timelike geodesics; the other two can be interpreted as the energy (per unit mass) and the projection of the angular momentum (per unit mass) along the rotation axis, respectively.

Inserting ansatz (\ref{eq:action}) in (\ref{eq:hamilton-jacobi}), one obtains relations among the functions $t(\tau)$, $r(\tau)$, $\theta(\tau)$ and $\phi(\tau)$; and, differentiating with respect to $\tau$, the following system of first-order, ordinary differential equations:
\begin{subequations}\begin{align}
        \Sigma \dv{t}{\tau} &= a(\mathcal{L}-a \mathcal{E}\sin^2 \theta) + \frac{\varrho^2 + a^2}{\Delta}[\mathcal{E}(\varrho^2 + a^2) - \mathcal{L}a],\\
        \Sigma \dv{r}{\tau} &= \pm \sqrt{\mathcal{R}},\\
        \Sigma \dv{\theta}{\tau} &= \pm \sqrt{\Theta},\\
        \Sigma \dv{\phi}{\tau} &= \frac{\mathcal{L}}{\sin^2 \theta} - a \mathcal{E} + \frac{a}{\Delta} [\mathcal{E}(\varrho^2 + a^2) - \mathcal{L}a],
\end{align}\end{subequations}
where 
\begin{align}
    \mathcal{R} &= [\mathcal{E} (\varrho^2 + a^2) -\mathcal{L}a]^2 - \Delta[\mu^2 \varrho^2 + (\mathcal{L} -a \mathcal{E})^2 + \mathcal{Q}],\label{eq:eom1}\\
    \Theta &= \mathcal{Q}- \cos^2 \theta \bigg[ a^2 (\mu^2-\mathcal{E}^2) + \frac{\mathcal{L}^2}{\sin^2 \theta} \bigg],
    \label{eq:eom2}
\end{align}
Here $\mathcal{Q}$ is the Carter constant already hinted to in section \ref{subsec:ec_obs}. Its existence derives from the Killing tensor (\ref{eq:ktens}) via the constant 
\begin{equation}
\mathcal{K}:= K_{\mu \nu}\dv{x^\mu}{\tau} \dv{x^\nu}{\tau},
\end{equation}
as $\mathcal{Q}:= \mathcal{K}-(a \mathcal{E} - \mathcal{L})^2$. Its expression in this spacetime coincides with its Kerr homonym's:
\begin{equation}
    \mathcal{Q} = u_\theta^2 + \cos^2\theta \left[a^2 (1-\mathcal{E})^2 - \frac{\mathcal{L}}{\sin^2 \theta} \right].
\end{equation}

Note that the system (\ref{eq:eom1}--\ref{eq:eom2}) looks almost identical to its Kerr analogue: indeed, the right-hand sides are precisely those one would find performing the same analysis in a Kerr spacetime, charted by the Boyer--Lindquist coordinates $(t,\ \varrho,\ \theta,\ \phi)$.

Hence, one might expect that our metric and Kerr's share the same geodesics. Namely, given a Kerr geodesic $\big(t(\tau),\ \varrho(\tau),\ \theta(\tau),\ \phi(\tau)\big)$, one could guess that the curve $\big(t(\tau),\ r(\tau) = \sqrt{\varrho(\tau)^2 - \ell^2},\ \theta(\tau),\ \phi(\tau)\big)$ might be a geodesic of our spacetime (charted by $r$ as Boyer--Lindquist-like radius)  ---  at least as long as $\varrho(\tau) \geq \ell$. 

This however is not true, in general. Clearly, the reason is that the relation between $r$ and $\varrho$ is not a mere shift:
\begin{equation}
    \dv{r}{\tau} = \dv{\varrho}{\tau} \dv{r}{\varrho} = \dv{\varrho}{\tau} \frac{\varrho}{r}
\end{equation}
and therefore
\begin{equation}
    \Sigma \dv{\varrho}{\tau} = \pm \sqrt{\mathcal{R}} \quad \text{(Kerr geodesic)} \quad\nRightarrow\quad  \Sigma \dv{r}{\tau} = \pm \sqrt{\mathcal{R}} \quad \text{(rotating SV geodesic)}.
\end{equation}
Rather, using the Kerr-like coordinate $\varrho$, test particles in our spacetime feel a distorted effective potential in the radial direction:
\begin{equation}
    \Sigma \dv{\varrho}{\tau} = \pm \sqrt{1-\frac{\ell^2}{\varrho^2}} \sqrt{\mathcal{R}}.
\end{equation}

Note however that, analogously to what happens in Kerr, circular orbits do exist on the equator $\theta=\pi/2$ with $\mathcal{Q} = 0$  ---  hence our analysis in section~\ref{subsec:orbits} is justified.


\bibliographystyle{JHEP.bst}
\bibliography{references}

\end{document}